\documentclass[10pt,a4paper]{article}
\usepackage{amsmath,latexsym,graphicx,psfrag,graphicx,caption2}
\usepackage{dsfont,cite}
\begin{document}
\headsep0cm
\topmargin0cm
\headheight0cm
\textheight24cm
\renewcommand{\d}{\mathrm{d}}
\newcommand{\arctanh}{\mathrm{arctanh}}
\newcommand{\arccot}{\mathrm{arccot}}
\newcommand{\arccsc}{\mathrm{arccsc}}
\newcommand{\sgn}{\mathrm{sgn}}
\newcommand{\prim}[1]{{#1^{\prime}}}
\newcommand{\tr}{\mathrm{Tr}}
\newcommand{\naeher}{\!\!\!}
\newcommand{\vect}[1]{\vec #1}
\newcommand{\ddd}[1]{\d\vect{#1} }

\begin{titlepage} 
\vspace*{-1cm} 
\begin{flushright} 
ZU--TH 14/05\\
hep-ph/0507152\\
July 2005
\end{flushright} 
\vskip 3cm 

\begin{center} 
{\Large\bf {\tt HPL},\vspace{0.1cm}\\a Mathematica implementation of the \vspace{0.1cm}\\harmonic polylogarithms}
\vskip 1.5cm  
{\large D.~Ma\^{\i}tre}
\vskip 1.5cm 
{\it Institut f\"ur Theoretische Physik, Universit\"at Z\"urich,
Winterthurerstrasse 190,\\ CH-8057 Z\"urich, Switzerland} 
\vskip 1.5cm 
\end{center}
\begin{abstract}
In this paper, we present an implementation of the harmonic polylogarithm of Remiddi and Vermaseren \cite{Remiddi} for Mathematica. It contains an implementation of the product algebra, the derivative properties, series expansion and numerical evaluation. The analytic continuation has been treated carefully, allowing the user to keep the control over the definition of the sign of the imaginary parts. Many options enables the user to adapt the behavior of the package to his specific problem.  
\end{abstract}
\vfill 
\end{titlepage} 

\section*{Package summary}
\begin{description}
\item[Title of the Package] HPL
\item[Version] 1.0
\item[Package obtained from] {\verb+http://www-theorie.physik.unizh.ch/~maitreda/HPL/+}
\item[E-Mail] maitreda@physik.unizh.ch
\item[License] none
\item[Computers] Computers running Mathematica under Windows or Linux
\item[Operating system] Linux or Windows
\item[Program language] Mathematica
\item[Memory required to execute] depending on the libraries loaded
\item[Other Package needed] none
\item[External file required] all needed files are provided  
\item[Keywords] Harmonic polylogarithm, multiple zeta values
\end{description}
\clearpage
\section{Introduction}
The harmonic polylogarithms (HPL's)\cite{Remiddi} are a generalization of the usual polylogarithms \cite{Lewin} and of the Nielsen polylogarithms \cite{Nielsen}. The HPL's appear in many calculations in high energy physics. They are found in three-loop deep inelastic splitting and coefficient functions \cite{Vermaseren:2005qc,Moch:2004xu,Vogt:2004mw,Moch:2004pa}, in two-loop massive vertex form factors \cite{Bonciani:2003te,Bonciani:2003ai,Bonciani:2003hc,Bernreuther:2004ih,Bernreuther:2004th,Bernreuther:2005rw,Mastrolia:2003yz,Degrassi:2005mc}, in two-loop Bhabha scattering \cite{Bonciani:2003cj,Bonciani:2004gi,Bonciani:2004qt,Czakon:2004wm,Bonciani:2005im,Penin:2005kf} and in multi-loop three-point and four-point functions \cite{Bern:2005iz,Birthwright:2004kk,Aglietti:2004ki,Heinrich:2004iq,Aglietti:2004nj,Aglietti:2004tq,Smirnov:2003vi,Aglietti:2003yc,Gehrmann:2001ck,Gehrmann:2000zt,Gehrmann:2002zr}. The HPL's also show up the expansion of hypergeometric functions around their parameters \cite{Weinzierl:2002hv,weinzierl1,Weinzierl:2004bn,Huber:2005yg} and in more formal developments \cite{Blumlein:2005jg}.

The HPL's have already been implemented for the algebraic manipulation language FORM \cite{FORM}. This implementation is described in \cite{Vermaseren_harmpol} and can be found at \cite{harmpol_homepage}. A FORTRAN code for the numerical evaluation of the HPL's is also available \cite{Gehrmann:2001pz}. A numerical implementation is also available for GiNaC \cite{Bauer:2000cp}, described in \cite{Vollinga:2004sn}. The aim of this work is to provide an implementation of the HPL's, including numerical evaluation. A {\tt Mathematica} implementation of the HPL's will in particular allow the usage of these functions in connection with many of the multi-purpose features of {\tt Mathematica}. Concerning speed, it is not expected that our {\tt Mathematica} implementation outperforms the FORM implementation\cite{Vermaseren_harmpol}. 

The paper is articulated as follows. In section \ref{definition} we review the definition of the HPL's. Their analytic properties are described in section \ref{derivative} (derivative), \ref{algebra} (product algebra) and  \ref{divergences} (singular behavior). In section \ref{basis} we treat the different identities relating different HPL's. Section \ref{series} presents the series expansion of the HPL's. The HPL's of related arguments are treated in section \ref{related}. Section \ref{analytic} treats the analytic continuation of the HPL's. Values of HPL of argument 1 are related to the multiple zeta values (MZV), which we discuss in section \ref{MZV}. In section \ref{implementation} we describe our implementation and provide examples of its usage. Finally we conclude in section \ref{conclusion}. Some tables and calculational details have been relegated to the appendix. In appendix \ref{MZVtable} we list some identities for the MZV. In appendix \ref{zetaproof} we prove the relation between the HPL's of unity argument and the MZV's. Finally in appendix \ref{identities} we list the representation of some HPL's in terms of more common functions.   
\section{Definition}\label{definition}
The harmonic polylogarithms (HPL) $H\left(a_1,\dots,a_k;x\right)$ are functions of one variable $x$ labeled by a vector $a=(a_1,\dots,a_k)$. The dimension $k$ of the vector $a$ is called the weight of the HPL. We define the functions 
\begin{eqnarray}
f_1(x)&=&\frac{1}{1-x}\nonumber\\
f_0(x)&=&\frac{1}{x}\nonumber\\
f_{-1}(x)&=&\frac{1}{1+x}
\end{eqnarray}
The HPL's are defined recursively through integration of these functions. For weight one we have
\begin{eqnarray}
H(1;x)&=&\int\limits_0^xf_1(t)\d t=\int\limits_0^x\frac{1}{1-t}\d t=-\log(1-x)\nonumber\\
H(0;x)&=&\log(x)\nonumber\\
H(-1;x)&=&\int\limits_0^xf_{-1}(t)\d t=\int\limits_0^x\frac{1}{1+t}\d t=\log(1+x),
\end{eqnarray}
and for higher weights 
\begin{eqnarray}
H(^n0;x)&=&\frac{1}{n!}\log^n x\nonumber\\
H(a,a_{1,\dots,k};x)&=&\int\limits_0^x f_a(t)H(a_{1,\dots,k};t)\d t\, ,
\end{eqnarray}
where we used the notations
\[^n i=\underbrace{i,\dots,i}_n,\quad \textnormal{and }\quad a_{1,\dots,k}=a_1,\dots,a_k .\]
A useful notation introduced in\cite{Remiddi} for harmonic polylogs with non-zero right-most index is given by dropping the zeros in the vector $a$, adding 1 to the absolute value of the next right non-zero index for each dropped 0. This gives for example $(3,-2)$ for $(0,0,1,0,-1)$. We can extend this notation to all index vectors by allowing zeros to take place in the extreme right of the new index vector. This gives for example $(3,-2,0,0)$  for $(0,0,1,0,-1,0,0)$. We will enclose index vectors in this notation in curly brackets and refer to it as the "m"-notation, as opposed to the "a"-notation. Some formulae or transformations are easier expressed in the one or the other notation, therefore we keep both notations in parallel.
\section{Derivatives}\label{derivative}
The formula for the derivative of the HPL's follows directly from their definition
\begin{equation}
\frac{\d}{\d x}H(a,a_{1,\dots,k};x)=f_a(x)H(a_{1,\dots,k};x).
\end{equation}
In the "m"-notation, the derivation reads for $n$ positive
\begin{eqnarray}
\frac{\d}{\d x}H\left(\{n,m_{1,\dots,k}\};x\right)&=&\frac{1}{x}H\left(\{n-1,m_{1,\dots,k}\};x\right),\qquad n>1\nonumber\\
\frac{\d}{\d x}H\left(\{-n,m_{1,\dots,k}\};x\right)&=&\frac{1}{x}H\left(\{-(n-1),m_{1,\dots,k}\};x\right),\qquad n>1\nonumber\\
\frac{\d}{\d x}H\left(\{1,m_{1,\dots,k}\};x\right)&=&\frac{1}{1-x}H\left(\{m_{1,\dots,k}\};x\right)\nonumber\\
\frac{\d}{\d x}H\left(\{-1,m_{1,\dots,k}\};x\right)&=&\frac{1}{1+x}H\left(\{m_{1,\dots,k}\};x\right)\nonumber\\
\frac{\d}{\d x}H\left(\{^n0\};x\right)&=&\frac{1}{x}H\left(\{^{n-1}0\};x\right)
\end{eqnarray}
\section{Product algebra}\label{algebra}
The product of two HPL of weights $w_1$ and $w_2$ can be expressed as a linear combination of HPL's of weight $w=w_1+w_2$. The formula in the "a"-notation for two HPL's with index vectors $\mathbf{p}$ and $\mathbf{q}$ is given by  
\[H(p_1,\dots,p_{w_1};x)H(q_1,\dots,q_{w_2};x)=H(\mathbf{p};x)H(\mathbf{q};x)=\sum\limits_{r\in \mathbf{p}\uplus\mathbf{q}}H(\mathbf{r};x)\]
where $\mathbf{p}\uplus\mathbf{q}$ is the set of all arrangements of the elements of $\mathbf{p}$ and $\mathbf{q}$ such that the internal order of the elements of $\mathbf{p}$ and $\mathbf{q}$ is kept. For example one has for $\mathbf{p}=(a,b)$ and $\mathbf{q}=(y,z)$
\begin{eqnarray}
H(a,b;x)H(y,z;x)&=&H(a,b,y,z;x)+H(a,y,b,z;x)\nonumber\\
&&+H(a,y,z,b;x)+H(y,a,b,z;x)\nonumber\\
&&+H(y,a,z,b;x)+H(y,z,a,b;x)
\end{eqnarray}
\section{Extraction of the singular behavior}\label{divergences}
The HPL's can have divergences in $x=0$ and $x=1$. The divergent part can be extracted with the help of the above product rules. Divergences in 0 appear as logarithmic divergence with behavior $\simeq \log^n(x)$ if there are $n$ 0's at the right end of the index vector. In this case one can make the divergent behavior explicit by writing 
\begin{eqnarray}
\lefteqn{H(a_{1,\dots,k};x)H(0;x)=}&&\nonumber\\
&&H(a_{1,\dots,k},0;x)+H(a_{1,\dots,k-1},0,a_k;x)+\dots+H(0,a_{1,\dots,k};x).
\end{eqnarray}
which one can solve for $H(a_{1,\dots,k},0;x)$. Recalling that  $H(0;x)=\log(x)$, the divergent $\log$ stands now explicitely. If there are more zeros in the right end of the index vector, one has to use this method recursively, until all end zeros have been exchanged against $\log$'s.

Similarly, when there are $n$ 1's at the very left of the index vector there appears a logarithmic divergence $\simeq\log^n(1-x)$. This can be extracted in the same way as above. One uses the product rule 
\begin{eqnarray}
\lefteqn{H(1;x)H(a_{1,\dots,k};x)=}&&\nonumber\\
&&H(1,a_{1,\dots,k};x)+H(a_1,1,a_{2,\dots,k};x)+\dots+H(a_{1,\dots,k},1;x).
\end{eqnarray}
Here we solve for $H(1,a_{1,\dots,k};x)$. Again we can apply the method recursively to make the singular behavior by $x=1$ explicit.

This method allows to express all HPL with left $1$'s and/or right $0$'s in terms of products of $H(1;x)$, $H(0;x)$ and HPL's without left $1$'s and right $0$'s. 

Using the same method, it is also possible to extract simultaneously the $H(-1;x)$ and $H(0;x)$ coming from the $-1$ on the left and $0$ on the right of the index vector. The factors $H(-1;x)$ will correspond to divergences in $-1$ which are relevant for the analytical continuation to the interval $(-\infty,-1)$, see section \ref{analytic}. One cannot however extract simultaneously the $H(-1;x)$ and  $H(1;x)$ by 
\[H(1,-1;x)=H(-1;x)H(1;x)-H(-1,1;x).\]
We will make extensive use of these relations in the next sections.
\section{Minimal set}\label{basis}
The procedure for extracting the divergent parts of an HPL described in section \ref{divergences} allows to express many HPL's in terms of HPL's without divergences at 0 and 1 (so called "irreducible") and products of HPL's of smaller weight. The product rules described in section \ref{algebra} also provide relations between HPL's of a given weight and HPL's of lower weight. One can combine all these relations to get a minimal set of HPL's for a given weight from which one can construct all other HPL's of this given weight, up to products of HPL's of lower weight. Table \ref{basistable} shows the number of HPL's, irreducible HPL's and the dimension of the minimal set as a function of the weight.  
\begin{table}[h]
\begin{center}
\begin{tabular}[h]{|c|ccc|}
\hline
\rule[-1.5ex]{0cm}{4ex}Weight&Full basis&Irreducible set&Minimal set\\\hline
1&3&3&3\\
2&9&4&3\\
3&27&12&8\\
4&81&36&18\\
5&243&108&48\\
6&729&324&116\\
7&2187&972&312\\
8&6561&2916&810\\\hline
\end{tabular}

\end{center}
\caption{Dimension of the different basis}
\label{basistable}
\end{table}\\
Only the number of elements in the minimal set is fixed, there is a freedom left for the choice of which elements are to be taken as independent. Our choice was first to exclude all HPL's whose divergent behavior can be extracted along the lines of section \ref{divergences} from the minimal set. For the remaining ones, we ordered the index vectors with the following procedure. One adds one to all indices of the index vector (in the "a"-notation), the result is to be interpreted as the expansion in basis 3 of a number. This number describs the index vector on a unique way. We used this numbering to sort the irreducible HPL's and choose to express the last as a function of the first.
\section{Series expansion}\label{series}
The series expansion of the HPL's can be defined recursively. Let us call $Z_i(a_{1,\dots,k})$ the coefficients of the expansion of $H\left(a_{1,\dots,k};x\right)$. 
\begin{equation}
H(a_{1,\dots,k};x)=\sum\limits_{i=0}^\infty x^i Z_i(a_{1,\dots,k})
\end{equation}
We assume that the index vector has no trailing 0, as these lead to $\log(x)$ divergences. These divergences have to be made explicit by the procedure of section \ref{divergences}. Now we use the definition of the HPL's
\[H(a,a_{1,\dots,k};x)=\int_0^x \d x'f_a(\prim{x})H(a_{1,\dots,k};\prim{x})=\sum\limits_{i=0}^\infty Z_i(a_{1,\dots,k})\int_0^x \d x'f_a(\prim{x})\prim{x}^i.\]
For the three different possibilities $1,0,-1$ for $a$ we get
\begin{eqnarray}
\int_0^x \frac{\d \prim{x}}{1-\prim{x}}\prim{x}^i&=&\int_0^x\d \prim{x}\sum\limits_{j=i}^{\infty}x^j=\sum\limits_{j=i+1}^\infty\frac{x^j}{j}\nonumber\\
\int_0^x \frac{\d \prim{x}}{\prim{x}}\prim{x}^i&=&\frac{x^i}{i}\nonumber\\
\int_0^x \frac{\d \prim{x}}{1+\prim{x}}\prim{x}^i&=&\!\!\int_0^x\d \prim{x}(x)^{i}\left(\sum\limits_{j=1}^{\infty}(-x)^j\right)=(-1)^{i+1}\sum\limits_{j=i+1}^\infty\frac{(-x)^j}{j}.
\end{eqnarray} 
The recursion relation for the $Z_i$'s is found by interchanging the order of the summations over $i$ and $j$
\begin{eqnarray}
Z_j(1,a_{1,\dots,k})&=&\frac{1}{j}\sum\limits_{i=2}^j Z_{i-1}(a_{1,\dots,k})\nonumber\\
Z_j(0,a_{1,\dots,k})&=&\frac{1}{j} Z_{j}(a_{1,\dots,k})\nonumber\\
Z_j(-1,a_{1,\dots,k})&=&\frac{(-1)^{j}}{j}\sum\limits_{i=2}^j(-1)^{i+1}Z_{i-1}(a_{1,\dots,k})
\end{eqnarray}
For the "m"-notation, we can use the same notation for the coefficient of the series expansion, as no confusion is possible.  
\[H\left(\{m_{1,\dots,k}\};x\right)=\sum\limits_{j=0}^\infty x^i Z_j(m_{1,\dots,k})\]
Here again we assume that all $\log(x)$ divergences have been made explicit by the procedure of section \ref{divergences}, so that the vector $m_{1,\dots,k}$ has no trailing 0. The recursion relations for $n$ positive reads
\begin{eqnarray}\label{coeffseries}
Z_j(n,m_{1,\dots,k})&=&\frac{1}{j^n}\sum\limits_{i=2}^j Z_{i-1}(m_{1,\dots,k})\nonumber\\
Z_j(-n,m_{1,\dots,k})&=&\frac{(-1)^j}{j^n}\sum\limits_{i=2}^j(-1)^{i+1}Z_{i-1}(m_{1,\dots,k}).
\end{eqnarray}
The recursion is settled by the expansions of $H(\{n\};x)$ and $H(\{-n\};x)$
\begin{eqnarray}
H(\{n\};x)&=&Li_n(x)=\sum\limits_{i=1}^\infty\frac{x^i}{i^n}\nonumber\\
H(\{-n\};x)&=&-Li_{n}(-x)=\sum\limits_{i=1}^\infty\frac{-(-x)^i}{i^n},
\end{eqnarray}
which give
\[Z_i(\{n\})=\left\{\begin{array}{cc}\displaystyle\frac{\sgn(n)^{i+1}}{i^{|n|}},\qquad& i>0\\0,&i\le 0\; .\end{array}\right.\]
\section{HPL's of related arguments}\label{related}
In this section, we present the identities between HPL's of different (but related) arguments. All the identities are obtained through change of variable in the definition of the HPL's. The easiest case is the change of variables 
\[x=-t.\]
 This transformation is only allowed if the right-most index is not zero, as in this case one would get negative argument in a logarithm\footnote{See section \ref{analytic} for the analytical continuation in this case}.
\begin{equation}\label{minustrans}
H\left(\{m_1,\dots,m_k\},-x\right)=(-1)^kH\left(\{-m_1,\dots,-m_k\};x\right)
\end{equation}
This identity holds in the "m"-notation. For index vectors in the "a"-notation, the exponent $k$ is not the length of the vector, but the number of $1$ and $-1$ in the index vector. 

We now consider the change of variable 
\[x\rightarrow x^2.\]
Since we cannot express $1+x^2$ as sum of our basis functions $f_1,f_0$ and $f_{-1}$, we will exclude index vectors with negative indices for our considerations. The identities for the weight $1$ are
\begin{eqnarray}
H(0;x^2)&=&\log(x^2)=2H(0;x),\nonumber\\
H(1;x^2)&=&\log(1-x^2)=H(1;x)-H(-1;x).
\end{eqnarray}
There we chose $x$ to be the positive root of $x^2$. Furthermore we see in the second equation that the identity only holds for $x$ smaller than 1. For higher weights we use recursively the relations 
\begin{eqnarray}
H(0,m_{2\dots,k};x^2)&=&\int_0^{x^2}\frac{\d \prim{x}}{\prim{x}}H(m_{2,\dots,k};\prim{x})\nonumber\\
&=&2\int_0^{x}\frac{\d \prim{t}}{\prim{t}}H(m_{2,\dots,k};\prim{t}^2),\nonumber\\
H(1,m_{2\dots,k};x^2)&=&\int_0^{x^2}\frac{\d \prim{x}}{1-\prim{x}}H(m_{2,\dots,k};\prim{x})\nonumber\\
&=&\int_0^{x}\d \prim{t}\left(\frac{1}{1-\prim{t}}-\frac{1}{1+\prim{t}}\right)H(m_{2,\dots,k};\prim{t}^2).
\end{eqnarray}
where $H(m_{2,\dots,k};\prim{t}^2)$ is expressed as HPL's of argument $\prim{t}$, which is known in a recursive approach.

The next transformation we consider is 
\[x\rightarrow 1-x.\]
Since $1/(2-x)$ (the transform of $1/(1+x)=f_{-1}$) can not be expressed as linear combinations of the $f_i$'s, we will only consider index vectors without negative index. We first have the identities for weight 1
\begin{eqnarray}
H(0;1-x)&=&\log(1-x)=-H(1;x),\nonumber\\
H(1;1-x)&=&-\log(x)=-H(0;x).
\end{eqnarray}
These identities only hold for $x$ between 0 and 1. Here again, we process recursively in the depth. We use the fact that one can express HPL's with 1's on the left of the index vector as product of $H(1;x)$ (whose transformation we know from above) and HPL's without left 1's and treat only the case of a 0 as the left index. For this we use the formula
\begin{eqnarray}
H(0,m_{2\dots,k};1-x)&=&\int_0^{1-x}\frac{\d \prim{x}}{\prim{x}}H(m_{2,\dots,k};\prim{x})\nonumber\\
&=&\int_0^{1}\frac{\d \prim{t}}{\prim{t}}H(m_{2,\dots,k};\prim{x})-\int_{1-x}^{1}\frac{\d \prim{x}}{\prim{x}}H(m_{2,\dots,k};\prim{x})\nonumber\\
&=&H(0,m_{1,\dots,k};1)-\int_0^{x}\frac{\d \prim{t}}{1-\prim{t}}H(m_{2,\dots,k};1-\prim{t}),\nonumber\\
\end{eqnarray}
where one has to insert for $H(m_{2,\dots,k};1-\prim{t})$ the expansion in terms of HPL's of argument $\prim{t}$ known from the recursion. The HPL's (with positive $m$'s) evaluated for argument 1 are related to the multiple zeta values (MZV). This aspect is treated in section \ref{MZV}. 

We consider now the transformation\footnote{The convention for the sign of the imaginary part of $x$ is different from \cite{Remiddi}.} 
\[x=\frac{1}{t} +i\epsilon\,, \qquad\qquad t=\frac{1}{x}-i\epsilon\,,\qquad\qquad x>1.\] 
The identities for weight 1 read
\begin{eqnarray}
H(0;x)&=&-H(0;t),\nonumber\\
H(1;x)&=&H(1;t)+H(0;t)+ i\pi,\nonumber\\
H(-1;x)&=&H(-1;t)-H(0;t).
\end{eqnarray}
For higher weight we proceed by induction. We will assume that the leftmost element in the vector index is $0$ or $-1$, since we can extract the one's with the procedure described in section \ref{divergences}.  
\begin{eqnarray}
H(a,a_{2,\dots,k};x)&=&\int\limits_0^x\d x'f_a(x')H(a_{2,\dots,k};t)\nonumber\\
&=&\left(\int\limits_0^1\d x'-\int\limits_x^1\d x'\right)f_a(x')H(a_{2,\dots,k};x')\nonumber\\
&=&H(a,a_{2,\dots,k};1)+\int\limits_t^1\frac{\d t'}{t'^2}f_a\left(\frac{1}{t'}\right)H\left(a_{2,\dots,k};\frac{1}{t'}\right)
\end{eqnarray} 
We now distinguish the two cases $a=0$ and $a=-1$,
\begin{eqnarray}
\int\limits_t^1\frac{\d t'}{t'^2}f_0\left(\frac{1}{t'}\right)&=&\int\d t'\frac{1}{t'}\nonumber\\
\int\limits_t^1\frac{\d t'}{t'^2}f_{-1}\left(\frac{1}{t'}\right)&=&\int\d t'\left(\frac{1}{t'}-\frac{1}{1+t'}\right) .
\end{eqnarray}
Inserting for 
\[H\left(a_{2,\dots,k};\frac{1}{t'}\right)\]
its representation in terms of HPL's of argument $t'$ (which is known in a recursive approach) one gets terms of  the kind  
\begin{equation}
\int\limits_t^1 \d t' f_{b}(t')H(b_{2,\dots,k};t')=H(b,b_{2,\dots,k};1)-H(b,b_{2,\dots,k};t).
\end{equation}
The term $H(b,b_{2,\dots,k};1)$ is a finite constant, as $b$ is either $-1$ or $0$ and not one.

The next transformation is 
\[x=t/(t-1).\] 
Since $\log((2t-1)/(t-1))$ (the transform of $\log(1+x)$) and $\log(t/(t-1))$ (the transform of $\log(x)$) can not be expressed in terms of $f_{1,0,-1}(t)$, we restrict the following consideration to HPL without trailing $0$'s and without negative index. We have only one identity of weight 1,
\[H(1;x)=-H(1;t)\; ,\]
which only holds for $x,t<1$. For the higher weights we use the recursion relations
\begin{eqnarray}
H(0,m_{2\dots,k};x)&=&\int_0^{x}\frac{\d \prim{x}}{\prim{x}}H(m_{2,\dots,k};\prim{x})\nonumber\\
&=&\int_0^{t}\d \prim{t}\left(-\frac{1}{1-\prim{t}}\right)H\left(m_{2,\dots,k};\frac{\prim{t}}{\prim{t}-1}\right),\nonumber\\
H(1,m_{2\dots,k};x)&=&\int_0^{x}\frac{\d \prim{x}}{\prim{x}}H(m_{2,\dots,k};\prim{x})\nonumber\\
&=&\int_0^{t}\d \prim{t}\left(\frac{1}{1-\prim{t}}+\frac{1}{\prim{t}}\right)H\left(m_{2,\dots,k};\frac{\prim{t}}{\prim{t}-1}\right) \,,
\end{eqnarray}
where the HPL in the integrand is replaced by its expansion in terms of HPL's of $t$, as known from the recursion.

The last transformation we consider is 
\[x=\frac{1-t}{1+t}\,.\]
Here we can also consider negative index in the index vector. The identities for weight 1 are
\begin{eqnarray}
H(0;x)&=&-H(1;t)-H(-1;t),\nonumber\\
H(1;x)&=&-H(0;t)-H(-1;1)+H(-1;t),\nonumber\\
H(-1;x)&=&-H(-1;t)+H(-1;1).
\end{eqnarray} 
For the identities of higher weight, one uses again the fact that HPL's with 1's on the left of the vector index can be reduced. The recursion relations for the two other cases are
\begin{eqnarray}
H(0,m_{2\dots,k};x)&=&\int_0^{x}\frac{\d \prim{x}}{\prim{x}}H(m_{2,\dots,k};\prim{x})\nonumber\\
&=&\int_0^{1}\frac{\d \prim{x}}{\prim{x}}H(m_{2,\dots,k};\prim{x})-\int_{x}^{1}\frac{\d \prim{x}}{\prim{x}}H(m_{2,\dots,k};\prim{x})\nonumber\\
&=&H(0,m_{2,\dots,k};1)\nonumber\\
&&-\int_0^{t}\d \prim{t}\left(\frac{1}{1-\prim{t}}+\frac{1}{1+\prim{t}}\right)H\left(m_{2,\dots,k};\frac{1-\prim{t}}{1+\prim{t}}\right),\nonumber\\ \\
H(-1,m_{2\dots,k};x)&=&\int_0^{x}\frac{\d \prim{x}}{\prim{x}}H(m_{2,\dots,k};\prim{x})\nonumber\\
&=&H(0,m_{2,\dots,k};1)-\int_0^{t}\frac{\d \prim{t}}{1+\prim{t}}H\left(m_{2,\dots,k};\frac{1-\prim{t}}{1+\prim{t}}\right),\nonumber\\
\end{eqnarray}
where we insert for \[H\left(m_{2,\dots,k};\displaystyle\frac{1-\prim{t}}{1+\prim{t}}\right)\]
 the identity of lower weight known from the recursion. 
\section{Analytical continuation}\label{analytic}
The transformations of the preceding section were valid for restricted intervals. In this section we consider the analytical continuation to the remaining real axis. We first consider the continuation to the interval $(-1,0)$. We define 
\[x=-t+\delta \epsilon i\;,\]
where $0<t<1$ and $\epsilon$ is infinitesimally small and positive. $\delta$ is either $+1$ or $-1$. The continuation for the HPL of weight 1 are
\begin{eqnarray}\label{acm10}
H(1;x)&=&H(-1;t)\,,\nonumber\\
H(0;x)&=&H(0;t)+\delta i\pi\,,\nonumber\\
H(-1;x)&=&H(1;t)\,.
\end{eqnarray} 
For higher weight one extracts the factors $H(0;x)$ with the method of section \ref{divergences} and substitutes the above result for it. For the remainder, one can use the formula (\ref{minustrans}) of the preceding section.

The next interval to consider is the interval $(1,\infty)$. There we use the transformation 
\[x\rightarrow t=\frac{1}{x}\,,\]
which transforms the real axis greater 1 into the interval $(0,1)$. We consider
\[x=\frac{1}{t} +i\delta \epsilon\,, \qquad\qquad t=\frac{1}{x}-i\delta\epsilon\,, \qquad\qquad x>1.\] 
This  transformation has already been treated in the preceding section for $\delta =1$. We can use the same procedure to keep track of the sign of the infinitesimal imaginary part by modifying the transformation of $H(1;x)$ to 
\[H(1;x)=H(1;t)+H(0;t)+ i\delta\pi\,.\]
The last interval is $(-\infty,-1)$. There one also uses the transformation $x\rightarrow -x$. Again we consider 
\begin{equation}
x=-t+\delta \epsilon i,\qquad t=-x-\delta \epsilon i.
\end{equation} The identities of weight 1 are
\begin{eqnarray}\label{acminfm1}
H(1;x)&=&-H(-1;t)=-H\left(-1;\frac{1}{t}\right)+H\left(0;\frac{1}{t}\right)\nonumber\\
H(0;x)&=&H(0;t)+\delta i\pi=-H\left(0;\frac{1}{t}\right)+\delta i\pi\nonumber\\
H(-1;x+\delta \epsilon i)&=&-H(1;t-\delta \epsilon i)=-H\left(1;\frac{1}{t}\right)-H\left(0;\frac{1}{t}\right)+\delta i\pi
\end{eqnarray}
For higher weight one extracts both the $H(0;x)$ and the $H(-1;x)$ with the method of section \ref{divergences} and substitutes the above expressions. The  remainder is transformed with (\ref{minustrans}). Since $t$ is larger than 1, one should use the analytical continuation described above to bring the argument of the HPL into the interval $(0,1)$.
\section{Values at unity and Multiple Zeta Values}\label{MZV}
Some transformations of the preceding section, there appear HPL's of argument $1$. These are related for positive $m$'s to the Multiple Zeta Values (MZV) and for general $m$'s to colored MZV. The relation can be found through induction\footnote{see Appendix \ref{zetaproof}.} and reads
\begin{eqnarray}\label{HPLMZVlink}
H(\{m_{1,\dots,k}\};1)&=&N(m_{1,\dots,k})\zeta(\tilde m_{1,\dots,k}),\quad k>1\\
H(\{m\};1)&=&\zeta(m)\,,\qquad\qquad \qquad \qquad m>0\\
H(\{-m\};1)&=&(1-2^{1-m})\zeta(m)\qquad\qquad m>0\,.
\end{eqnarray}
The MZV $\zeta$ are defined as
\begin{equation}\label{MZVdef}
\zeta(m_1,\dots,m_k)=\sum\limits_{i_1=1}^\infty\sum\limits_1^{i_1-1}\dots\sum\limits_1^{i_{k-1}-1}\prod\limits_{j=1}^k\frac{\sgn(m_j)^{i_j}}{i_j^{|m_j|}}\,.
\end{equation}
and are described in the literature, for example in \cite{math.CA/9910045,Borwein:1996yq}. The vector $\tilde m$ is obtained from the vector $m$ with
\begin{equation}
\tilde m=(m_1,\sgn(m_1)m_2,\dots, \sgn(m_{i-1}) m_i,\dots ,\sgn(m_{k-1}) m_k)
\end{equation}
The factor $N(m_{1,\dots,k})$ is given by
\begin{equation}
N(m_{1,\dots,k})=(-1)^{\#(m_i<0)}
\end{equation}
 The MZV's also form an algebra. Due to this fact, they can be expressed in terms of a few mathematical constants like powers of $\pi$, $\zeta$-functions and polylogs at specified values. We list some of the identities of \cite{math.CA/9910045,Borwein:1996yq} in Appendix \ref{MZVtable}.
For the implementation of the HPL at unity, we translated the tables of the FORM package {\tt harmpol.h} \cite{Remiddi} and their expansions for weight 7 and 8 {\tt htable7.prc} and {\tt htable8.prc} for Mathematica.  

In these tables, there appear some constants that are not expressible through known constants like $\pi$, $\zeta(n)$, $\log(2)$, or $Li_n(1/2)$. Using the different relations between the different MZV, one can reduce the number of independent constants. Which constants are kept is a matter of choice. In the tables we translated, the choice of the independent constants is
\begin{eqnarray}
s_6  &=& S(\{-5,-1\},\infty) =\sum\limits_{i_1=1}^{\infty}\frac{(-1)^{i_1}}{i_1^5} \sum\limits_{i_2=1}^{i_1}\frac{(-1)^{i_2}}{i_2}\nonumber\\
&=&\zeta(-5,-1)+\zeta(6)\simeq 0.98744142640329971377\\
s_{7a}  &=& S(\{-5,1,1\},\infty) =\sum\limits_{i_1=1}^{\infty}\frac{(-1)^{i_1}}{i_1^5} \sum\limits_{i_2=1}^{i_1}\frac{1}{i_2}\sum\limits_{i_3=1}^{i_2}\frac{1}{i_3}\nonumber\\
&=&\zeta(-5,1,1)+\zeta(-6,1)+\zeta(-5,2)+\zeta(-7)\\
&\simeq& -0.95296007575629860341\\
s_{7b}  &=& S(\{5,-1,-1\},\infty) =\sum\limits_{i_1=1}^{\infty}\frac{1}{i_1^5} \sum\limits_{i_2=1}^{i_1}\frac{(-1)^{i_2}}{i_2}\sum\limits_{i_3=1}^{i_2}\frac{(-1)^{i_3}}{i_3}\nonumber\\
&=&\zeta(7)+\zeta(5,2)+\zeta(-6,-1)+\zeta(5,-1,-1)\\
&\simeq&  1.02912126296432453422\nonumber\\
s_{8a}  &=& S(\{5,3\},\infty) =\sum\limits_{i_1=1}^{\infty}\frac{1}{i_1^5} \sum\limits_{i_2=1}^{i_1}\frac{1}{i_2^3}\nonumber\\
&=&\zeta(8)+\zeta(5,3)\simeq 1.0417850291827918834\nonumber\\
s_{8b}  &=& S(\{-7,-1\},\infty) =\sum\limits_{i_1=1}^{\infty}\frac{(-1)^{i_1}}{i_1^7} \sum\limits_{i_2=1}^{i_1}\frac{(-1)^{i_2}}{i_2}\nonumber\\
&=&\zeta(8)+\zeta(-7,-1)\simeq 0.99644774839783766600\\
s_{8c}  &=& S(\{-5,-1,-1,-1\},\infty) \nonumber\\&=&\sum\limits_{i_1=1}^{\infty}\frac{(-1)^{i_1}}{i_1^5} \sum\limits_{i_2=1}^{i_1}\frac{(-1)^{i_2}}{i_2}\sum\limits_{i_3=1}^{i_2}\frac{(-1)^{i_3}}{i_3}\sum\limits_{i_3=1}^{i_3}\frac{(-1)^{i_4}}{i_4}\nonumber\\
&=&\zeta(8)+\zeta(-7,-1)+\zeta(-5,-3)+\zeta(6,2)+\zeta(-5,-1,2)\nonumber\\
&&+\zeta(-5,2,-1)+\zeta(6,-1,-1)+\zeta(-5,-1,-1,-1)\\
&\simeq& 0.98396667382173367094\nonumber\\
s_{8d}  &=& S(\{-5,-1,1,1\},\infty) =\sum\limits_{i_1=1}^{\infty}\frac{(-1)^{i_1}}{i_1^5} \sum\limits_{i_2=1}^{i_1}\frac{(-1)^{i_2}}{i_2}\sum\limits_{i_3=1}^{i_2}\frac{1}{i_3}\sum\limits_{i_3=1}^{i_3}\frac{i}{i_4}\nonumber\\
&=&\zeta(8)+\zeta(-5,-3)+\zeta(6,2)+\zeta(7,1)+\zeta(-5,-2,1)+\zeta(-5,-1,2)\nonumber\\
&&+\zeta(6,1,1)+\zeta(-5,-1,1,1)\simeq 0.99996261346268344768
\end{eqnarray}

For index vectors with left 1's, there appear divergences of the form
\[\sum\limits_{i=1}^{\infty}\frac{1}{i}=S(1,\infty)=H(1;1).\]
These divergences are well defined\footnote{see \cite{Remiddi}.} and can cancel during a calculation. Right 0's can be extracted with the method of section \ref{divergences}. All factors $H(0;1)=\log(1)=0$ vanish, even multiplied with $H(1;1)=-\log(0)$, since
\[ \log(x)\log^n(1-x)\rightarrow 0,\qquad x \rightarrow 0,\quad n>0.\]  
\section{Implementation}\label{implementation}
In this section we present our implementation of the HPL's. The package can be found at\cite{HPLhomepage}. Installation instructions can be found there. After instalation, the package can be called with

{\tt <<HPL`.}

\noindent This call should be done at the beginning of the {\tt Mathematica} session. 
\subsection{New functions}
In the package {\tt HPL} we defined the following new functions
\begin{itemize}
\item {\tt HPL[m,x]} is the harmonic polylogarithm $H(m;x)$. {\tt m} is a list representing the index vector. We chose the "m"-notation as the standard notation. It is possible to give as argument a vector in the "a"-notation, or even a mix between the two notations, the result will be automatically converted to the "m"-notation. 
\vspace{0.2cm}\\
\fbox{\parbox{0.75\textwidth}{\includegraphics[scale=0.5]{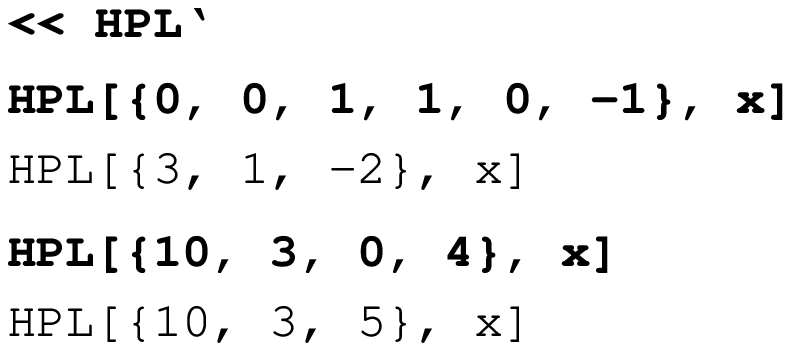}}
}\vspace{0.1cm}\\
\item {\tt HPLMtoA[m\_List]},{\tt HPLAtoM[a\_List]} convert vectors of the "m"- to the "a"- and from "a"- to "m"-notation respectively. Both can convert vectors which mix the two notations.
\vspace{0.2cm}\\
\fbox{\parbox{0.75\textwidth}{\includegraphics[scale=0.5]{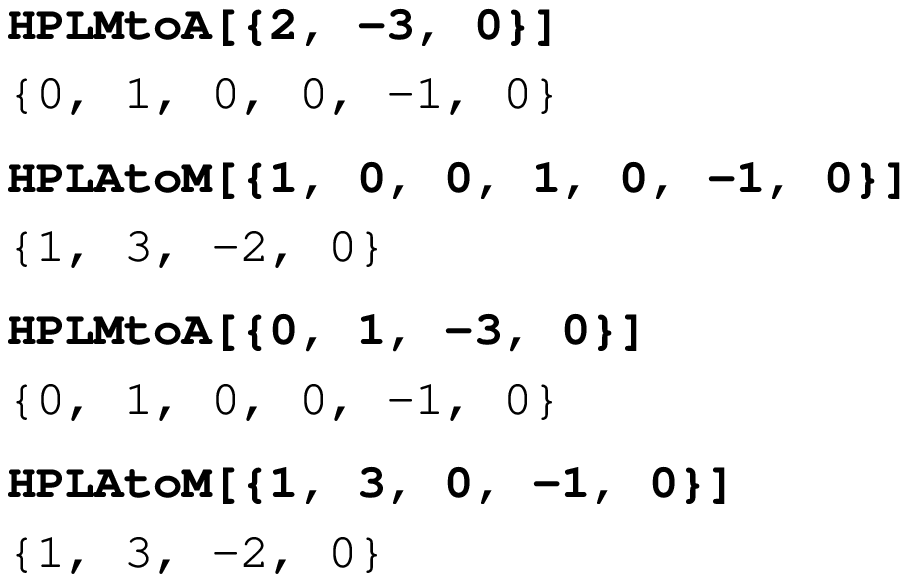}}
}\vspace{0.1cm}\\
\item {\tt HPLLogExtract} extracts the logarithmic divergences of the HPL in its argument at 0 and 1. The result is displayed as function of $\log(x),\log(1-x)$ or $H(1;x),H(0;x)$ depending on the option settings (see section \ref{options}).
\vspace{0.2cm}\\
\fbox{\parbox{0.75\textwidth}{\includegraphics[scale=0.5]{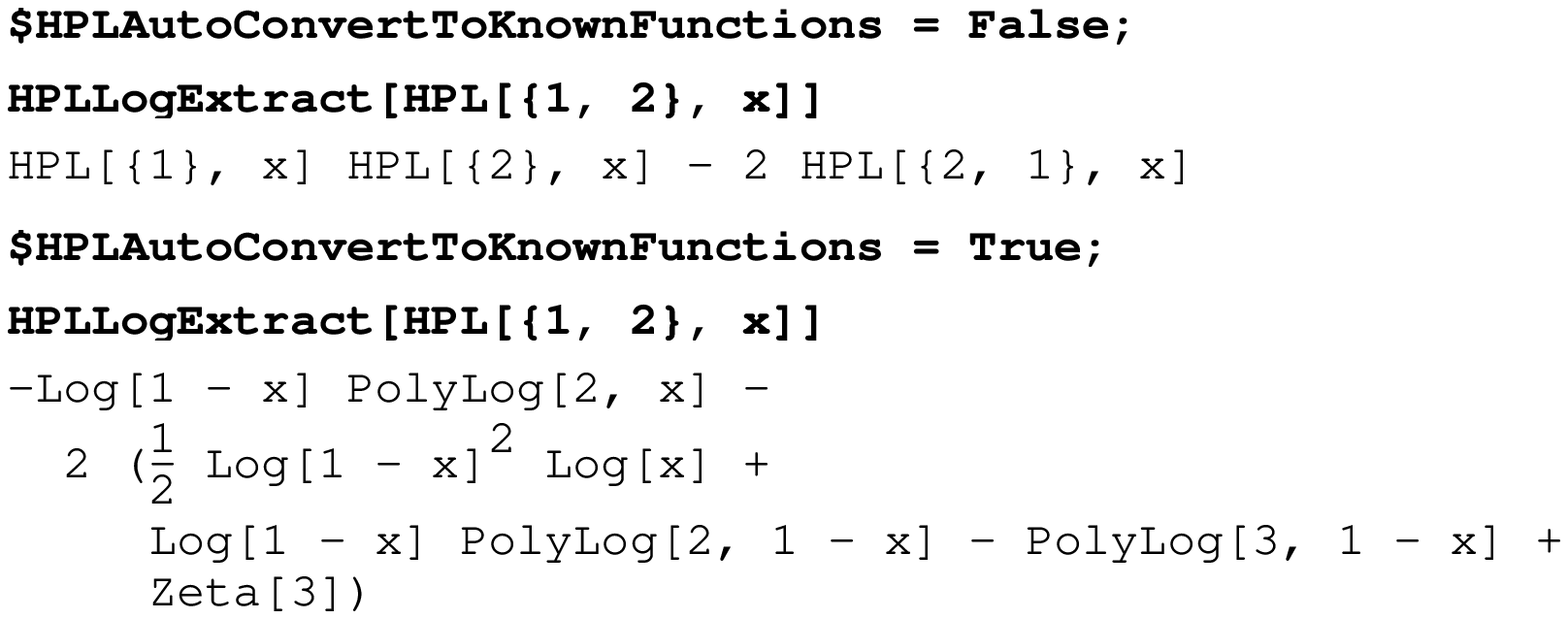}}
}\vspace{0.1cm}\\
\item {\tt HPLConvertToKnownFunctions} returns its argument with HPL's replaced by their representation in terms of more common functions, whenever possible. This is only needed if the option {\tt \$HPLAutoConvertToKnownFunctions} is set {\tt False} (see section \ref{options}).
\vspace{0.2cm}\\
\fbox{\parbox{0.75\textwidth}{\includegraphics[scale=0.5]{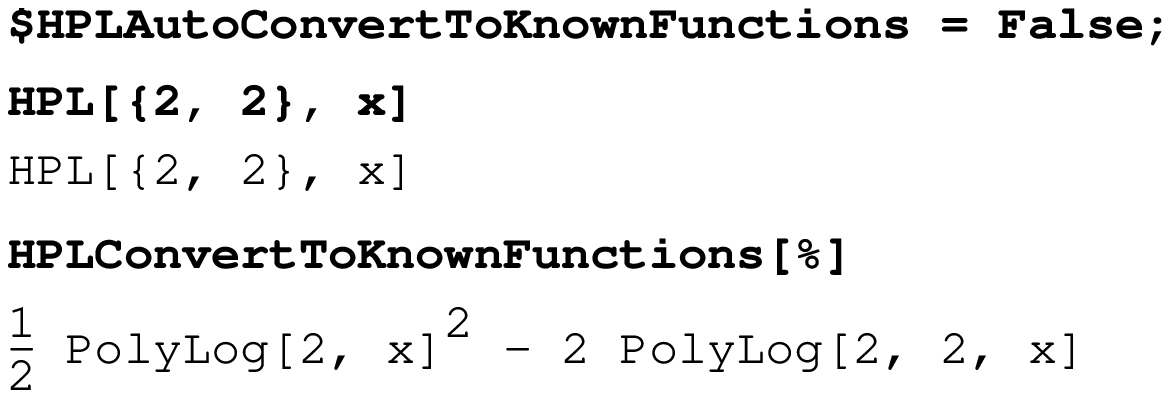}}
}\vspace{0.1cm}\\
\item {\tt HPLProductExpand} returns its argument where the products of HPL's of weight $w_1,\dots,w_k$ are replaced by their representation as a linear combination of HPL's of weight $w_1+\dots+w_k$. In order to expand all products, {\tt HPLProductExpand} expands the argument (using {\tt Expand}), so that terms of the form 
\[H(\dots;x)\big(H(\dots;x)+H(\dots;x)+\dots\big).\]
are also replaced.
\vspace{0.2cm}\\
\fbox{\parbox{0.75\textwidth}{\includegraphics[scale=0.5]{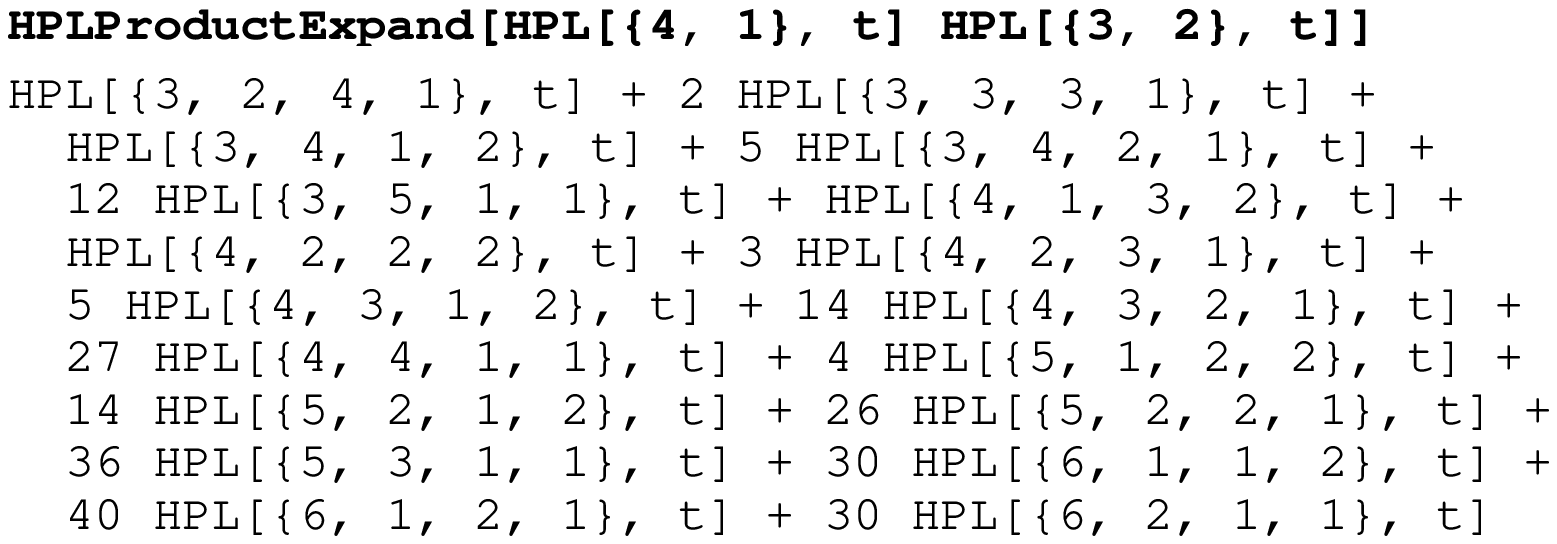}}
}\vspace{0.1cm}
\item {\tt HPLConvertToSimplerArgument} returns its argument with HPL's of arguments $-x$, $x^2$, $1-x$, $x/(x-1)$ and $(1-x)/(1+x)$ replaced by their expansion as a sum of HPL's of argument $x$, as described in section \ref{related}. 
\vspace{0.2cm}\\
\fbox{\parbox{0.75\textwidth}{\includegraphics[scale=0.5]{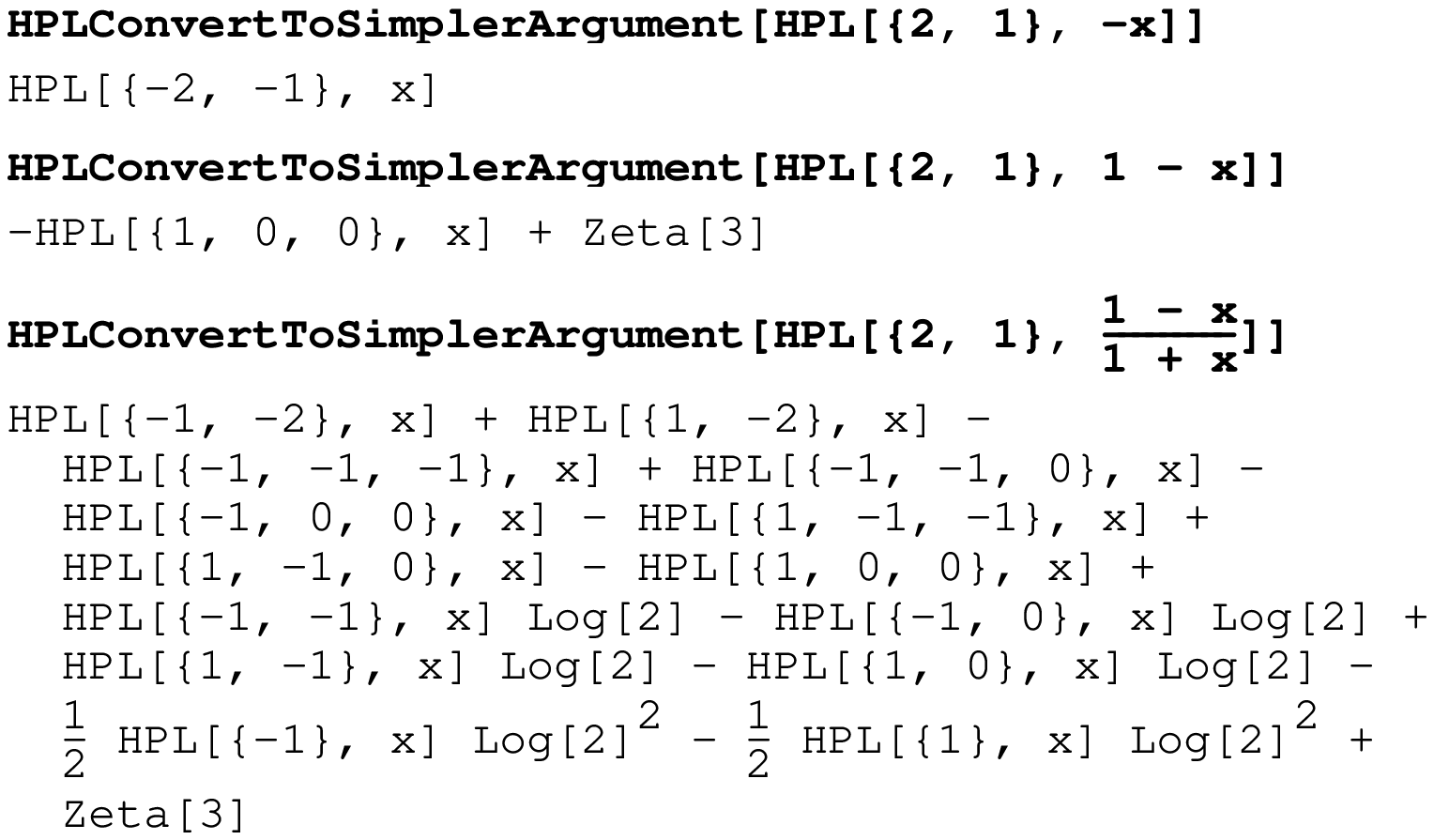}}
}\vspace{0.1cm}
\item {\tt HPLReduceToMinimalSet} returns its argument with the HPL's projected to the minimal set, as described in section \ref{basis}. 
\vspace{0.2cm}\\
\fbox{\parbox{0.75\textwidth}{\includegraphics[scale=0.5]{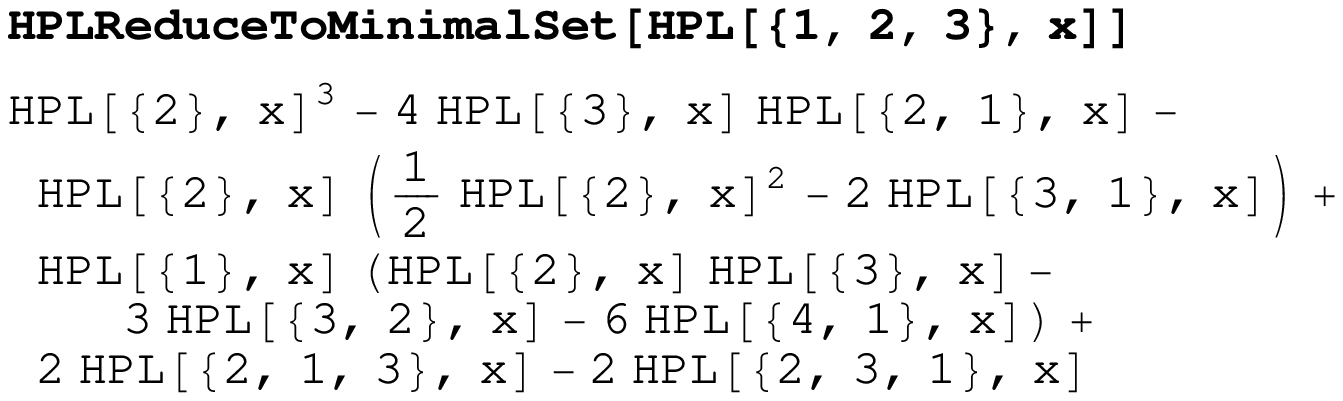}}
}\vspace{0.1cm}
\item {\tt HPLAnalyticContinuation} returns its argument with HPL's replaced by their analytic continuation. The arguments of the HPL's are taken to belong to the interval specified by the option  {\tt HPLAnalyticContinuationRegion} which can be either
\begin{description}
\item[{\tt minftom1}] the interval $-\infty$ to $-1$
\item[{\tt m1to0}] the interval $-1$ to $0$
\item[{\tt onetoinf}] the interval $1$ to $\infty$.
\end{description}
The HPL's are replaced by their representation in terms of HPL's of argument in the interval $0,1$. The choice of the side of the branch cut from which the argument is approached is set by the option 

{\tt HPLAnalyticContinuationSign} 

which can take values $-1$, $1$ or any symbol.
\vspace{0.2cm}\\
\fbox{\parbox{0.75\textwidth}{\includegraphics[scale=0.5]{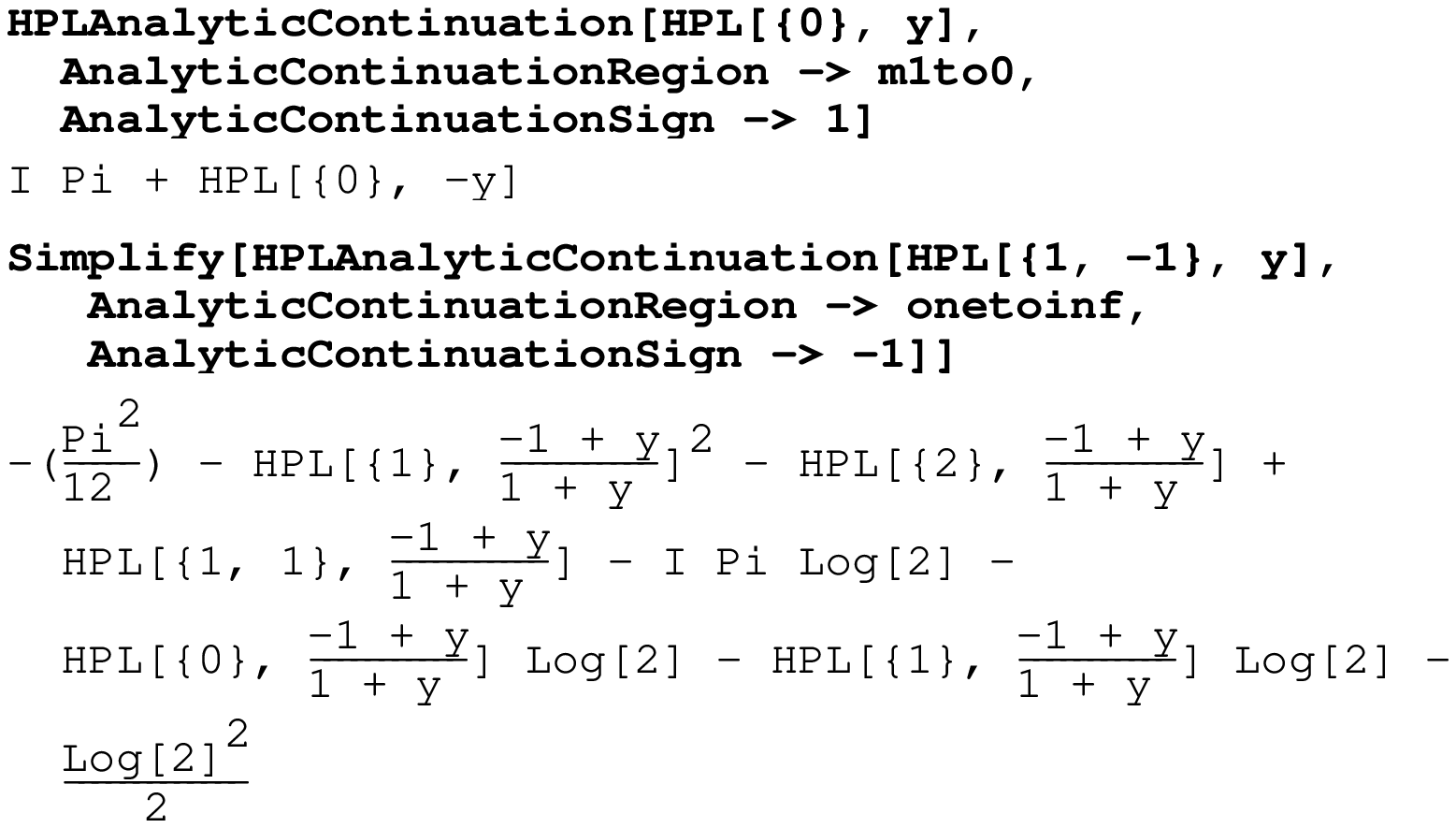}}
}\vspace{0.1cm}\\
If the option {\tt HPLAnalyticContinuationRegion} is omitted, and if the argument is numerical, {\tt HPLAnalyticContinuation} will use automatically the appropriate setting. If the option {\tt HPLAnalyticContinuationSign} is omitted, {\tt HPLAnalyticContinuation} will use the value stored in the variable {\tt \$HPLAnalyticContinuationSign} which is set by default to $1$\footnote{This is the same convention as \cite{Gehrmann:2001pz}, but opposite to that of \cite{Vollinga:2004sn}}. 
\vspace{0.2cm}\\
\fbox{\parbox{0.75\textwidth}{\includegraphics[scale=0.5]{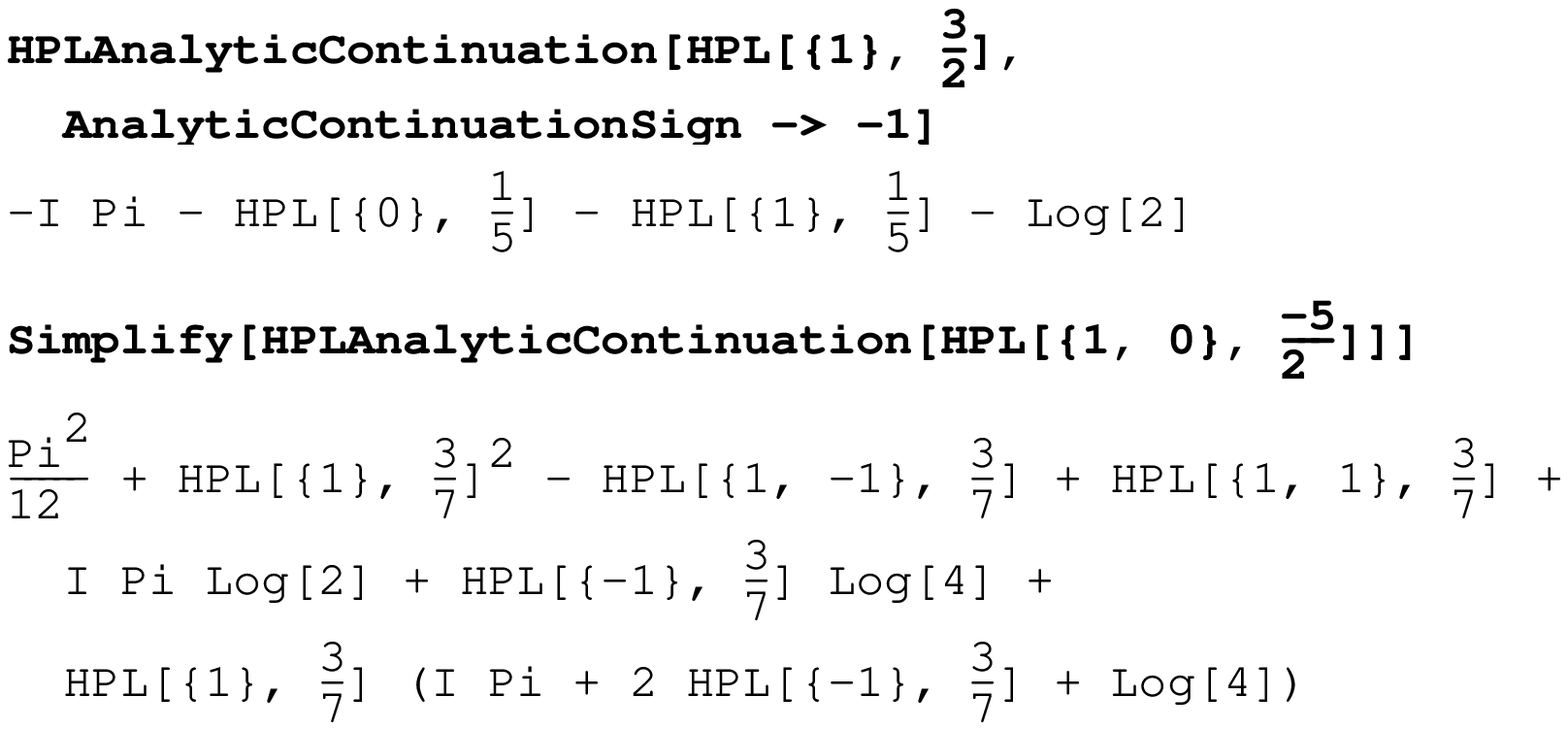}}
}\vspace{0.1cm}\\
It is to be noted that the {\tt Mathematica} conventions for the analytic continuation are not always the same as that of the {\tt HPL} package. This is illustrated by the following example
\vspace{0.2cm}\\
\fbox{\parbox{0.75\textwidth}{\includegraphics[scale=0.5]{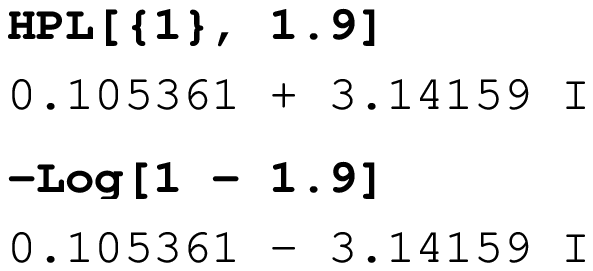}}
}\vspace{0.1cm}\\
Since the substitution of the HPL's through more common functions has precedence over the analytic continuation, the option 

{\tt \$HPLAutoConvertToKnownFunctions} 

can interfere with the analytic continuation, as shown in the following example.
\vspace{0.2cm}\\
\fbox{\parbox{0.75\textwidth}{\includegraphics[scale=0.5]{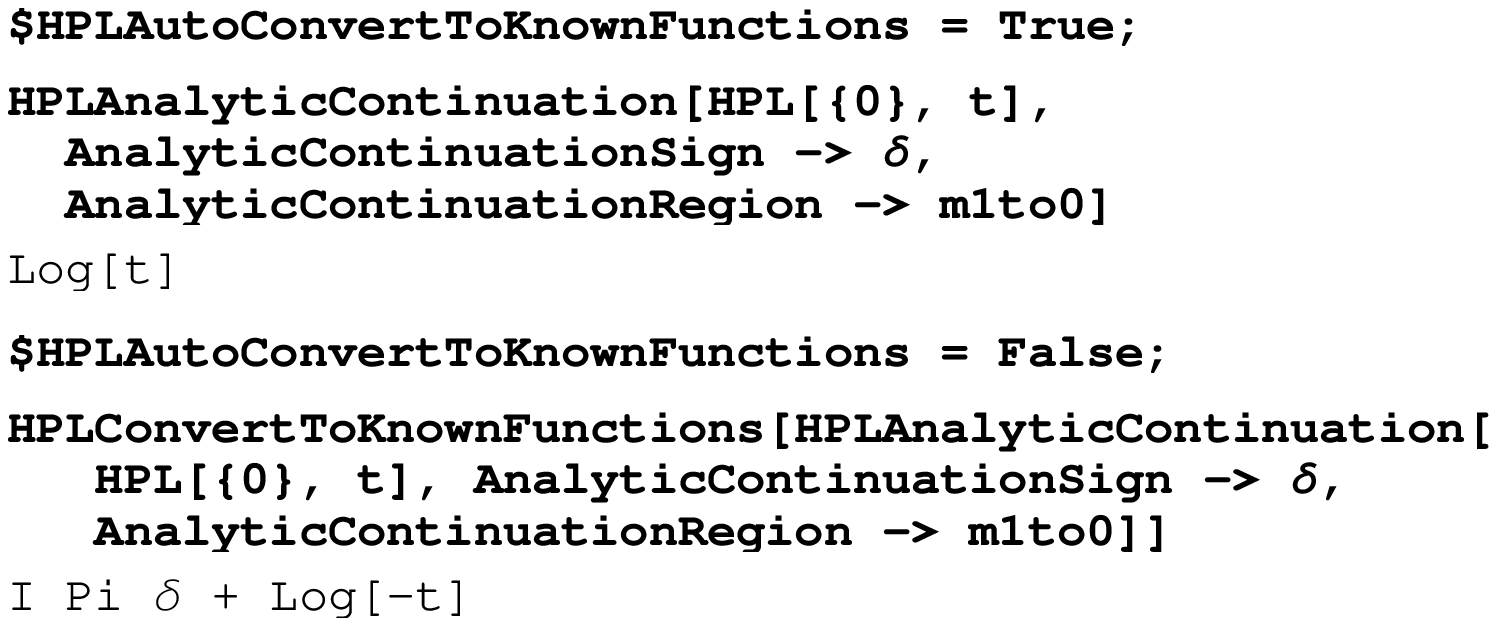}}
}\vspace{0.1cm}\\
This example shows that with the option 

{\tt \$HPLAutoConvertToKnownFunctions} 

set {\tt True} we loose the control over the sign of the imaginary part (as in the first case it will now depend on the {\tt Mathematica} conventions).
\item {\tt MZV[m]} is the Multiple Zeta Value (see section \ref{MZV}) corresponding to the index vector $m$. Their value in terms of mathematical constants are tabulated\footnote{these tables are those of the FORM package {\tt harmpol}\cite{Vermaseren_harmpol}} up to weight 8 and systematically replaced. For higher weights, the cases covered by Appendix \ref{MZVtable} are also replaced.
\vspace{0.2cm}\\
\fbox{\parbox{0.75\textwidth}{\includegraphics[scale=0.5]{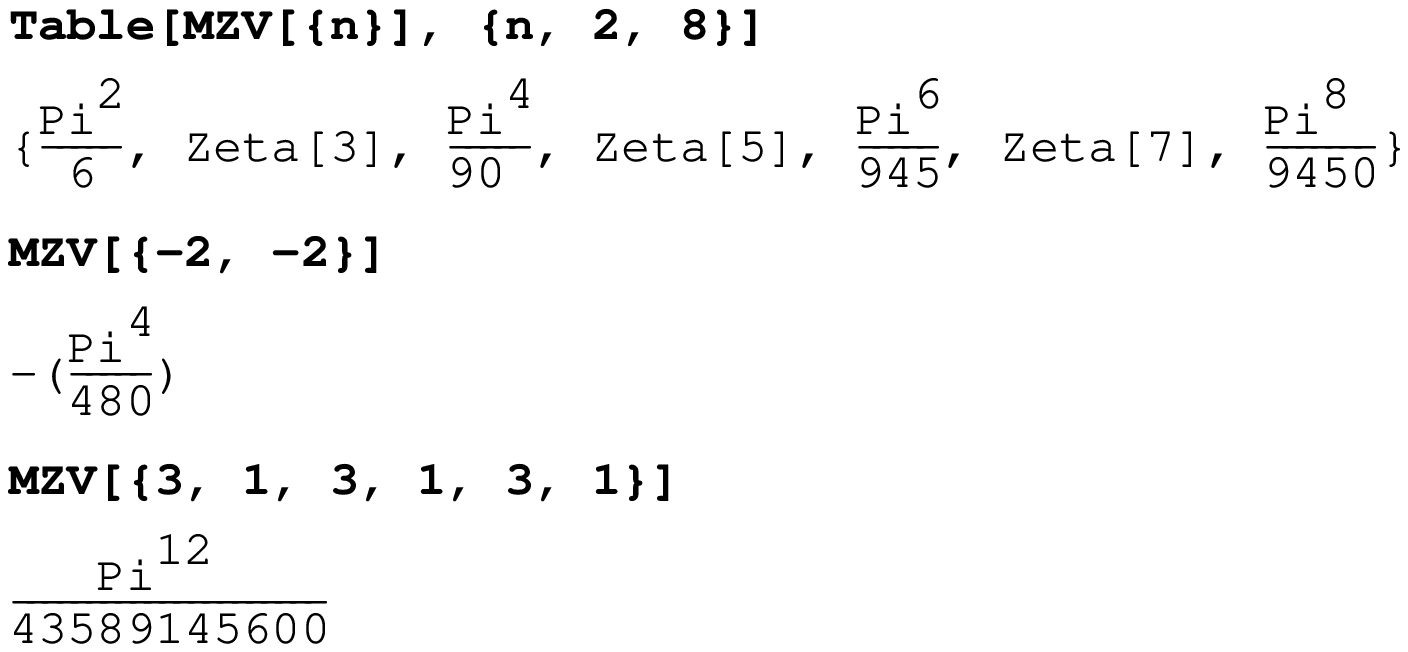}}
}\vspace{0.1cm}
\item The function {\tt \$HPLOptions} gives a list of the options of the package and their current values.
\vspace{0.2cm}\\
\fbox{\parbox{0.75\textwidth}{\includegraphics[scale=0.5]{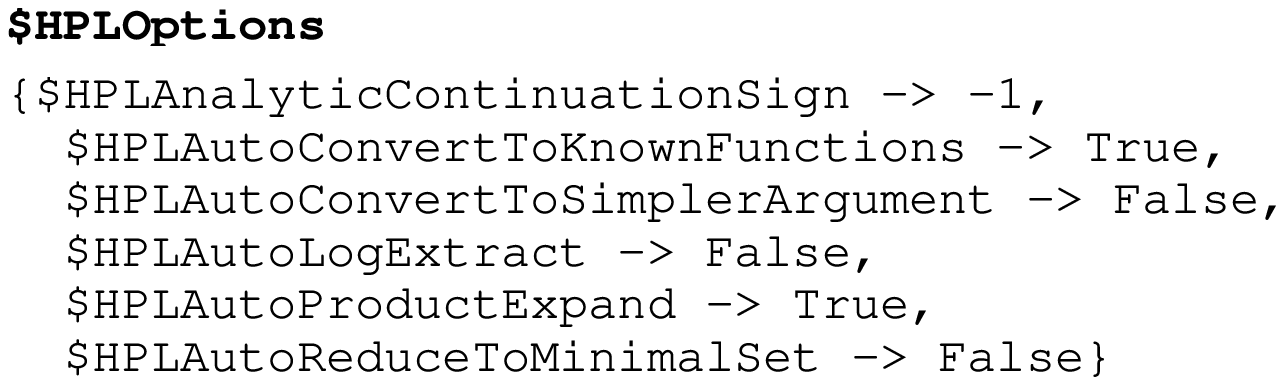}}
}
\end{itemize}
\subsection{Functions modified}
\begin{itemize}
\item We defined the derivatives of the HPL's as described in section \ref{derivative}. The integration showing up in the recursive definition of the HPL's is also implemented.
\vspace{0.2cm}\\
\fbox{\parbox{0.75\textwidth}{\includegraphics[scale=0.5]{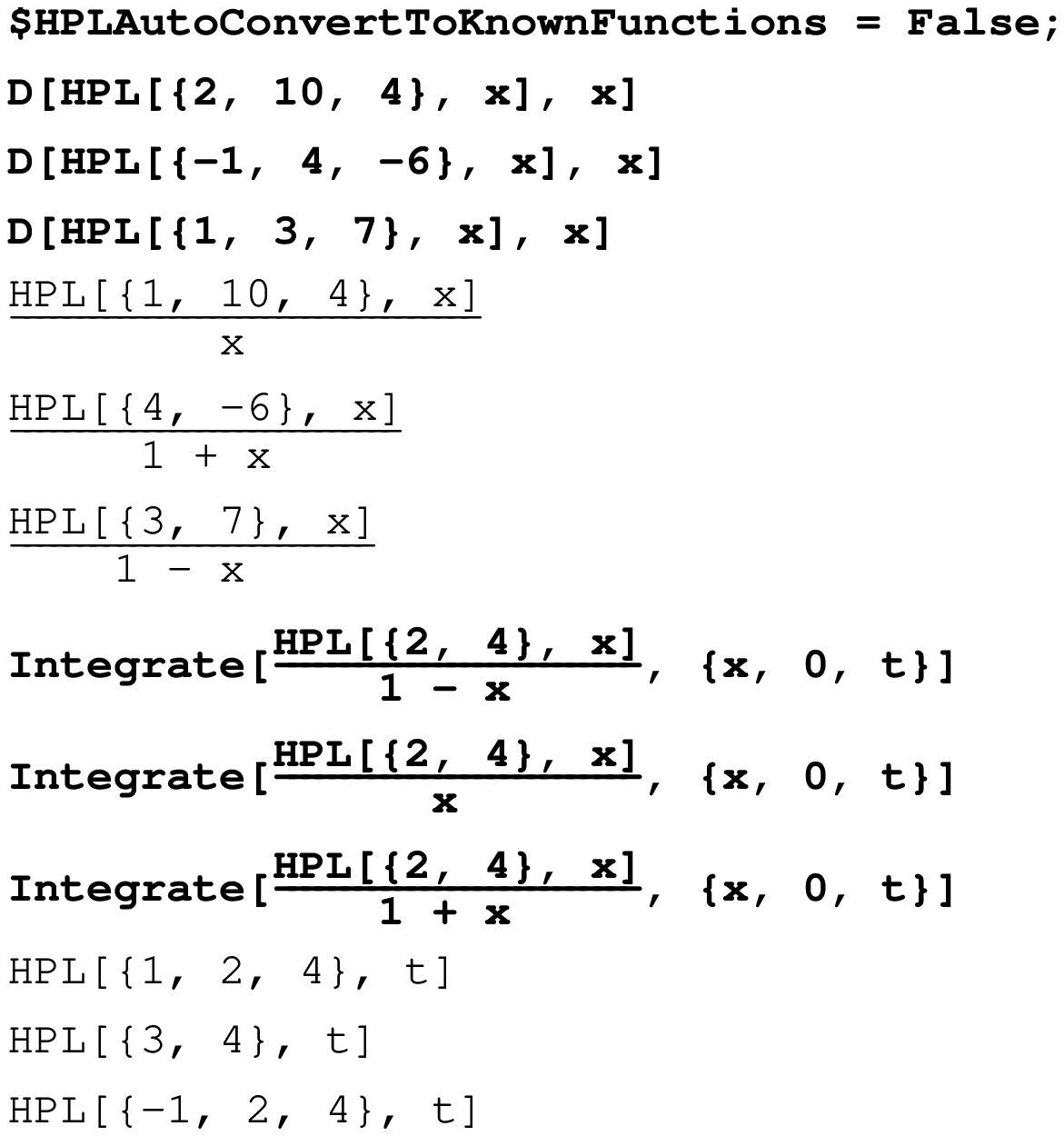}}
}\vspace{0.1cm}
\item The function {\tt Series} is able to expand HPL's around $x=0$ and $x=1$.
\vspace{0.2cm}\\
\fbox{\parbox{0.75\textwidth}{\includegraphics[scale=0.5]{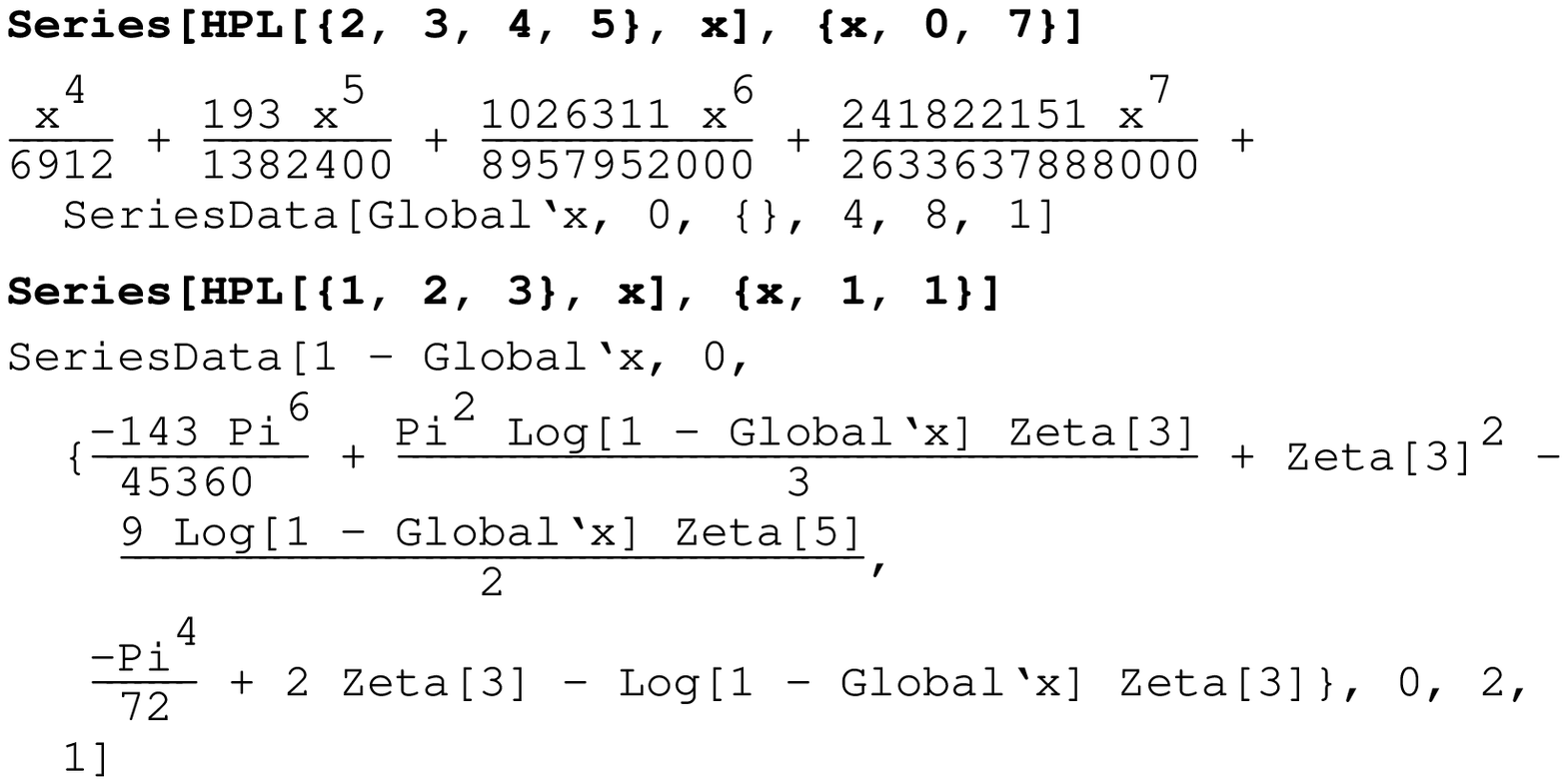}}
}\end{itemize}
\subsection{Working with options}\label{options}
The package {\tt HPL} has some options to control its behavior. They set the preferred form in which expressions are displayed. The option can be locally overridden by the functions described above. The effects of the options are described in the following.
\begin{description}
\item[{\tt \$HPLAutoLogExtract}:] If {\tt True} the logarithmic divergences  $\log(1-x)$ and $\log(x)$ are automatically extracted from the HPL's following the procedure described in section \ref{divergences}. These divergences are displayed as $\log$ or as $H(0/1;x)$, depending on the value of the option {\tt \$HPLAutoConvertToKnownFunctions}. The default setting is {\tt False}.  
\item[{\tt \$HPLAutoProductExpand}:] If {\tt True} the products of HPL's are automatically converted into a sum of HPL of weight equal to the sum of the weights of the two factors, as described in section \ref{algebra}. Default is {\tt False}.  
\item[{\tt \$HPLAutoConvertToKnownFunctions}:] If set {\tt True}, the HPL's will be converted to more common functions (logs, polylogarithms, polynoms, Nielsen polylogs) if possible, using the identities of Appendix \ref{identities}. This might be counterproductive when the properties of the HPL's are more explicit in the HPL form than in their equivalent representation. Furthermore, if this options is set {\tt True} while using the analytic continuation described above, the result may be wrong, as {\tt Mathematica} does not have different conventions for the analytic continuation. Default is {\tt False}.
\item[{\tt \$HPLAutoReduceToMinimalSet}:] If set {\tt True}, the HPL's will be automatically reduced to a minimal basis (up to weight 8). This only makes sense if one does not expand the obtained products again, or if the factors of smaller weight can be replaced by their expression in terms of known functions. Therefore, for the reduction to be performed, one has to have the option 

{\tt \$HPLAutoProductExpand} equal {\tt False} or 

{\tt \$HPLAutoConvertToKnownFunctions} equal {\tt True}. 

If this is not fulfilled, the option will have no effect. Default is {\tt False}.
\item[{\tt \$HPLAutoConvertToSimplerArgument}:] If set {\tt True}, the HPL's of arguments $-x$, $x^2$, $1-x$, $1/x$, $x/(x-1)$ and $(1-x)/(1+x)$ will be automatically substituted by their representation in term of HPL's of argument $x$ along the lines of section \ref{related}. Default is {\tt False}.
\item[{\tt \$HPLAnalyticContinuationSign}:] If set to 1 the analytic continuation of the HPL's is taken for arguments with an positive infinitesimal imaginary part, if set to -1 with an negative one. This is only the default setting and can be overridden by specifying the option {\tt AnalyticContinuationSign} in the function {\tt HPLAnalyticContinuation}, as described above. Default setting is $+1$.
\end{description}
\subsection{Numerical evaluation}
If the argument of the HPL is numerical, the package evaluates its value and gives the result in the same precision as the precision of $x$. 
In \cite{Remiddi}, the authors propose to use the argument transformation $x\rightarrow (1-t)/(1+t)$ for the evaluation of HPL's with argument near one or negative with a large absolute value. This is only of advantage if one wants to evaluate the whole set of HPL's for this given $x$. Since for low weight, we have a representation in terms of usual functions (which {\tt Mathematica} can evaluate fast and precisely), this change of variable is not useful in this case. For high weights, the expression of the HPL's in term of HPL's of argument $\frac{1-x}{1+x}$ is quite large, so that the time gain in the convergence is compensated by the number of different HPL's to evaluate. Therefore, we used the series expansion systematically for the evaluation of the numerical value of the HPL. The convergence is not very good for values near 1, where not all digits displayed are accurate. There is a FORTRAN package for the precise numerical evaluation of HPL's \cite{Gehrmann:2001pz}. 
\vspace{0.2cm}\\
\fbox{\parbox{0.75\textwidth}{\includegraphics[scale=0.5]{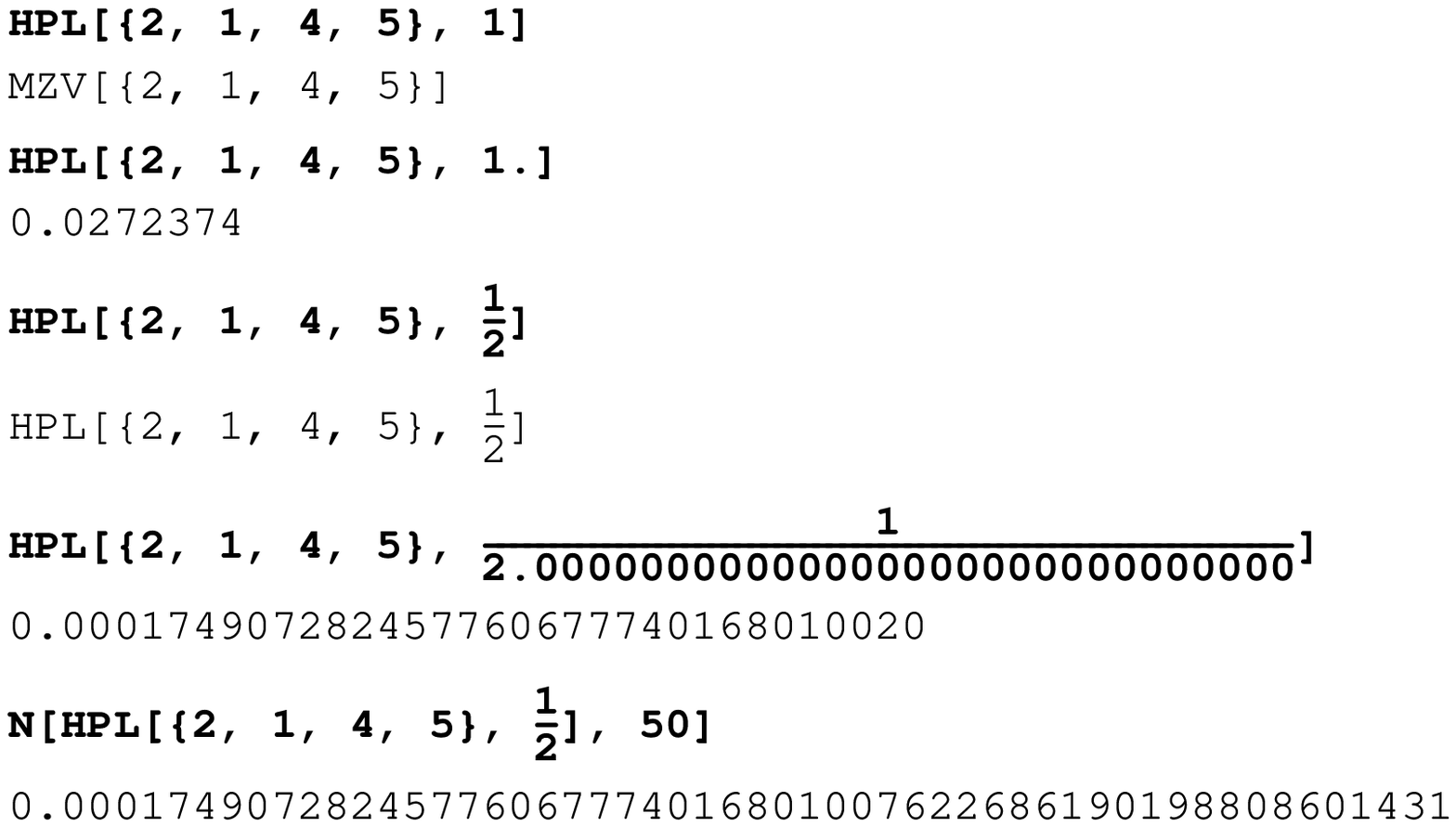}}
}\vspace{0.1cm}\\
For argument outside the interval $0,1$, one can give the sign of the infinitesimal imaginary part to be taken in account for the analytic continuation, as for the function {\tt HPLAnalyticContinuation}
\vspace{0.2cm}\\
\fbox{\parbox{0.75\textwidth}{\includegraphics[scale=0.5]{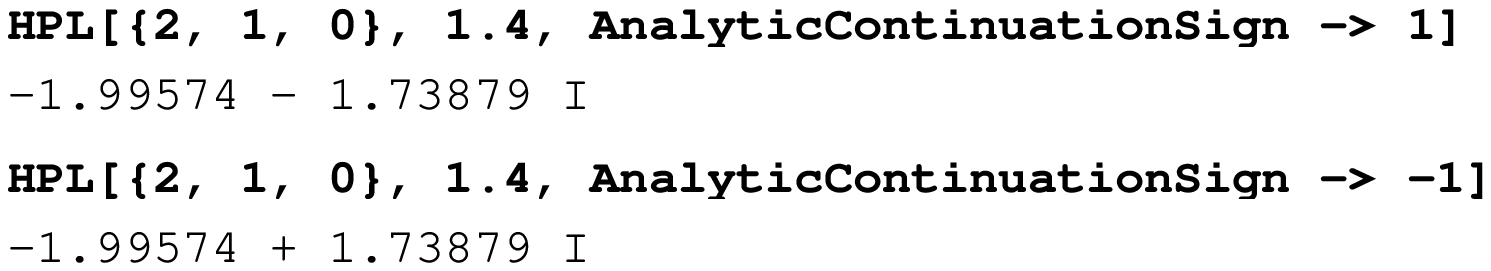}}
}\vspace{0.1cm}\\ 
For the numerical values of the MZV, we implemented the procedure described in \cite{math.CA/9910045,Vollinga:2004sn}. The numerical values of the constants $s_6,s_{7a},s_{7b},s_{8a},s_{8b},s_{8c}$ and $s_{8d}$ correspond to the results given by the {\tt EZface} application\cite{EZface}.
\vspace{0.2cm}\\
\fbox{\parbox{0.75\textwidth}{\includegraphics[scale=0.5]{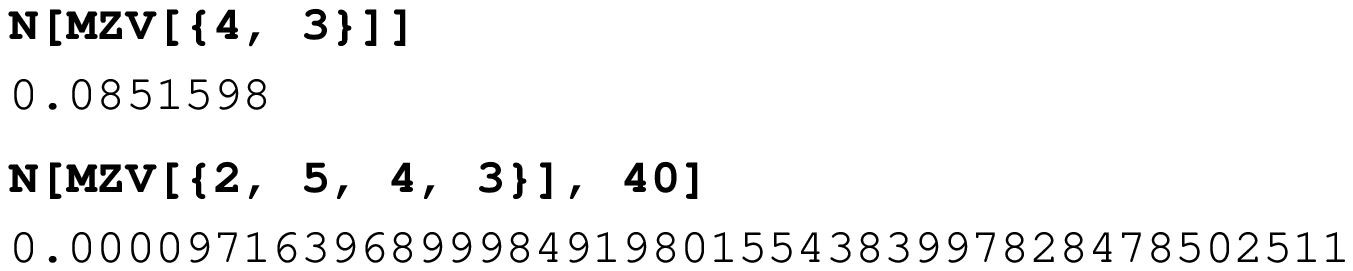}}}\vspace{0.2cm}\\
Due to the possibility of evaluating the HPL numerically, {\tt Plot} is able to represent HPL's
\vspace{0.2cm}\\
\fbox{\parbox{0.75\textwidth}{\includegraphics[scale=0.5]{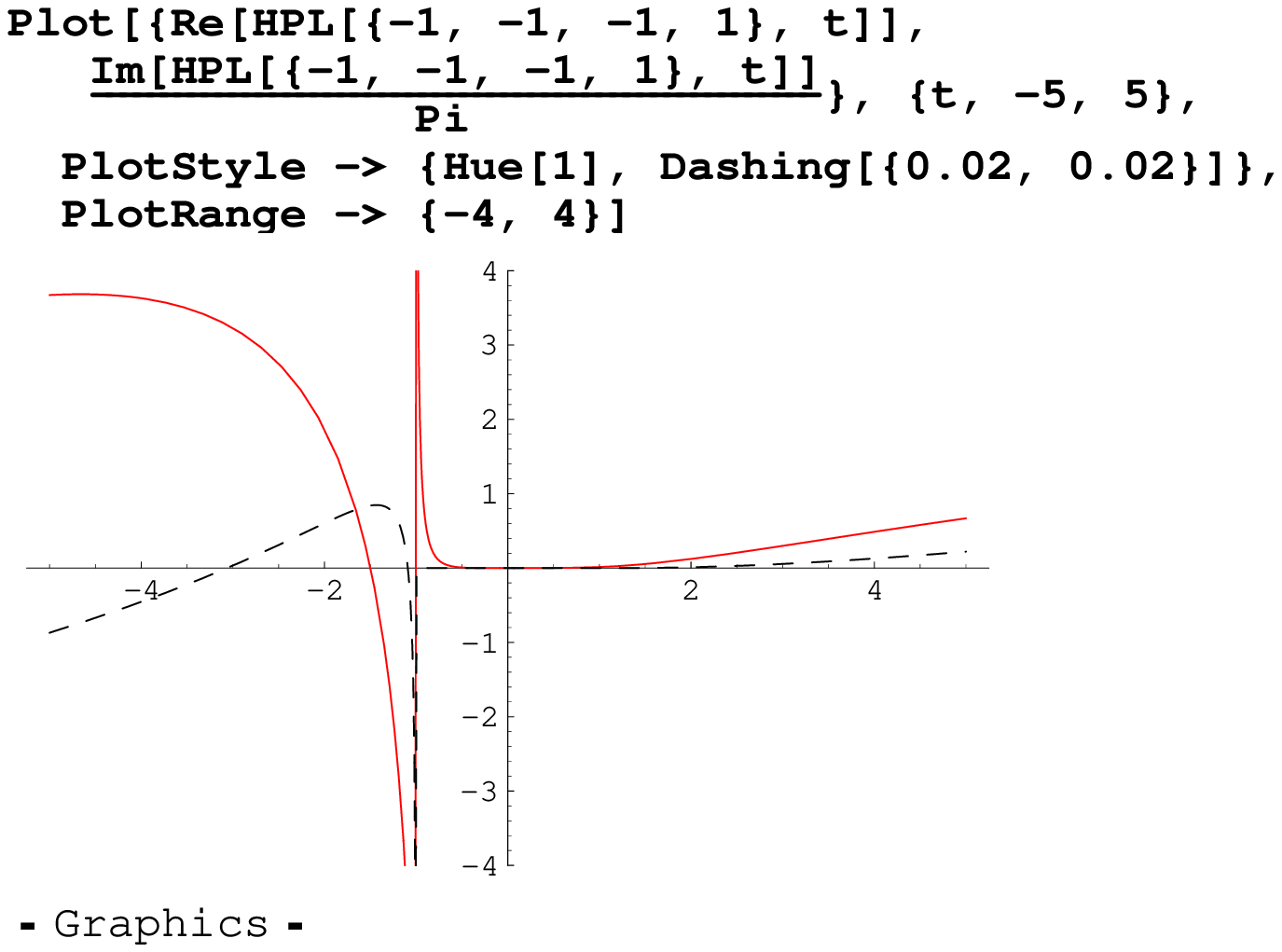}}
}\vspace{0.1cm}\\
We checked numerical agreement with \cite{Gehrmann:2001pz} at double precision accuracy.
\section{Conclusion}\label{conclusion}
In this paper, we presented an implementation of the harmonic polylogarithm of Remiddi and Vermaseren \cite{Remiddi} for Mathematica. It contains an implementation of the product algebra, the derivative properties, series expansion and numerical evaluation. The analytic continuation has been treated carefully, allowing the user the full control over the definition of the sign of the imaginary parts. Many options enable the user to adapt the behavior of the package to his specific problem. This package is already used by the package {\tt HypExp}\cite{Huber:2005yg} to treat the HPL's and MZV's appearing in the expansion of hypergeometric functions around integer-valued parameters.
\section*{Acknowledgement}
We would like to thank Thomas Gehrmann for many useful discussions and Jos Vermaseren, Ettore Remiddi, Thomas Gehrmann for carefully reading our manuscript. We also wish to thank the Swiss National Science Foundation (SNF) which supported this work under contract 200021-101874.
\appendix
\section{Multiple Zeta Values}\label{MZVtable}
 We list here some of the identities found in \cite{math.CA/9910045,Borwein:1996yq}.
\begin{eqnarray}
\zeta(2,1)&=&\zeta(3)\\
\zeta(4,2)&=&\zeta^2(3)-\frac{4\pi^6}{2835}\\
2 \zeta(m,1)&=&m\zeta(m+1)-\sum\limits_{k=1}^{m-2}\zeta(m-k) \zeta(k+1),\quad 2\le m \in \mathds{Z}\\
\zeta(2,\{1\}^n)&=&\zeta(n+2)\\
\zeta(3,\{1\}^n)&=&\zeta(n+2,1)\\
&=&\frac{n+2}{2}\zeta(n+3)-\frac{1}{2}\sum_{k=1}^n\zeta(k+1)\zeta(n+2-k)\\
\zeta(\{2\}^n)&=&\frac{2 (2 \pi)^{2n}}{(2n+1)!}\left(\frac{1}{2}\right)^{2n+1}\\
\zeta(\{4\}^n)&=&\frac{4 (2 \pi)^{4n}}{(4n+2)!}\left(\frac{1}{2}\right)^{2n+1}\\
\zeta(\{6\}^n)&=&\frac{6 (2 \pi)^{6n}}{(6n+3)!}\\
\zeta(\{8\}^n)&=&\frac{8 (2 \pi)^{8n}}{(8n+4)!}\left\{\left(1+\frac{1}{\sqrt{2}}\right)^{4n+2}+\left(1-\frac{1}{\sqrt{2}}\right)^{4n+2}\right\}\\
\zeta(\{3,1\}^n)&=&4^{-n}\zeta(\{4\}^n)=\frac{2\pi^{4n}}{(4n+2)!}\\
\zeta(2,\{1,3\}^n) &=&4^{-n}\sum_{k=0}^n(-1)^k\zeta(\{4\}^n)\big\{(4k+1)\zeta(4k+2)\nonumber\\
&&\quad\quad-4\sum_{j=1}^k\zeta(4j-1)\zeta(4k-4j_3)\big\}
\end{eqnarray}
\subsection{Colored MZV}
Here we list some identities for MZV with negative arguments found in \cite{Borwein:1996yq}
\begin{eqnarray}
\zeta(\{-2\}_n)&=&\frac{\pi^{2n}}{(2n+1)!}\frac{(-1)^{n(n+1)/2}}{2^n}\\
\zeta(\{-4\}_n)&=&\frac{\pi^{4n}}{(4n+2)!}\frac{(-1)^{n(n+1)/2}}{2^n}\left((1+\sqrt{2}^{2n+1}+(1-\sqrt{2}^{2n+1}))\right)\\
\zeta(\{-6\}_n)&=&\frac{\pi^{6n}}{(6n+3)!}\frac{3}{2}\Big(1+2^{3n+1}(-1)^{n(n+1)/2}\\\nonumber
&&\times\left\{\left(\frac{1+\sqrt{3}}{2}\right)^{6n+3}+\left(\frac{1-\sqrt{3}}{2}\right)^{6n+3}-1\right\}\\
\zeta(\{-1\}_n)&=&(-1)^n\sum\prod_{k\ge 1}\frac{1}{j_k!}\left(\frac{-Li_k((-1)^k)}{k}\right)^{j_k}
\end{eqnarray}
where the sum in the last equation is over all non negative integers satisfying $\sum_{k\ge 0}k j_k=n$
\begin{eqnarray}
\zeta(-1,\{1\}_n)&=&(-1)^{n+1}\frac{(\log2)^n}{n!}\\
\zeta(-1,-1,\{1\}_n)&=&-Li_{n+2}(1/2)
\end{eqnarray}
\section{Conversion of HPL at unity to MZV}\label{zetaproof}
We first prove 
\begin{eqnarray}
H(-m;x)&=&-\zeta(-m)=\sum_{i=1}^\infty\frac{(-1)^{i+1}}{i^m}\nonumber\\
&=&-\sum_{i=1}^\infty\frac{1}{(2i)^m}-\sum_{i=1}^\infty\frac{-1}{(2i-1)^m}\nonumber\\
&=&-2^{-m}\zeta(m)+(\sum_{i=1}^\infty\frac{1}{i^m}-\sum_{i=1}^\infty\frac{1}{(2i)^m}\nonumber\\
&=&-2^{-m}\zeta(m)+\left(\zeta(m)-2^{-m}\zeta(m)\right)\nonumber\\
&=&(1-2^{1-m})\zeta(m)
\end{eqnarray}
For the proof of the formula for higher depth we start with the definition (\ref{coeffseries}) of the coefficients of the series expansion of the HPL and first prove that 
\begin{equation}
(-1)^{i+1}Z_i(m_{1,\dots,k})=(-1)^{k+1}Z_i(-m_{1,\dots,k}).
\end{equation} 
We show that it holds for depth 1,
\begin{eqnarray}
\lefteqn{(-1)^{i+1}Z_i(m_1)=(-1)^{i+1}\frac{\sgn(m_1)^{i+1}}{i^{|m|}}}&&\nonumber\\
&&=\left\{\begin{array}{cc}
(-1)^{i+1}\displaystyle\frac{1}{i^{m_1}}=Z_i(-m_1),\qquad&m_1>0\\&\\(-1)^{i+1}\displaystyle\frac{(-1)^{i+1}}{i^{|m_1|}}=\frac{1}{i^{-m_1}}=Z_i(-m_1),&m_1<0.
\end{array}\right. 
\end{eqnarray} 
For $m_{1,\dots,k}$ with $m_1>0$ we get
\begin{eqnarray}
\lefteqn{(-1)^{i+1}Z_1(m_{1,\dots,k})=}&&\nonumber\\
&=&(-1)^{i+1}\frac{1}{i^{m_1}}\sum\limits_{i_2=1}^{i-1}Z_{i_2}(m_{2,\dots,k})\nonumber\\
&=&-\frac{(-1)^{i}}{i^{m_1}}\sum\limits_{i_2=1}^{i-1}(-1)^{i_2+1}(-1)^kZ_{i_2}(-m_{2,\dots,k})\nonumber\\
&=&(-1)^{k+1}Z_{i}(-m_{1,\dots,k}),
\end{eqnarray}
and for $m_1<0$,
\begin{eqnarray}
\lefteqn{(-1)^{i+1}Z_1(m_{1,\dots,k})=}&&\nonumber\\
&=&(-1)^{i+1}\frac{(-1)^{i+1}}{i^{|m_1|}}\sum\limits_{i_2=1}^{i-1}(-1)^{i_2+1}Z_{i_2}(m_{2,\dots,k})\nonumber\\
&=&\frac{(1}{i^{|m_1|}}\sum\limits_{i_2=1}^{i-1}(-1)^kZ_{i_2}(-m_{2,\dots,k})\nonumber\\
&=&(-1)^{k+1}Z_{i}(-m_{1,\dots,k}).
\end{eqnarray}
The next step is to prove for $k>1$
\begin{eqnarray}
\lefteqn{Z_i(m_{1,\dots,k})=}&&\nonumber\\
&&N(m_{1,\dots,k})\frac{\sgn(m_1)^i}{i^{|m_1|}}\sum\limits_{i_2=1}^{i-1}\dots\!\!\!\!\sum\limits_{i_k}^{i_{k-1}-1}\frac{\sgn( m_1m_2)^{i_2}}{i_2^{|m_2|}}\dots\frac{\sgn(m_{k-1}m_k)^{i_k}}{i_k^{|m_k|}}
\end{eqnarray}
 where $N(m_{1,\dots,k})$ is given by $(-1)^{\#n}$ with $\#n$ the number of negative indices in the index vector $m$.
  
The proof is again through induction in the depth. We first test the claim for depth 2. For $m_1$ positive
\begin{eqnarray}
\lefteqn{Z_i(m_1,m_2)=\frac{1}{i^{m_1}}\sum\limits_{i_2=1}^{i-1}Z_{i_2}(m_2)=\frac{1}{i^{m_1}}\sum\limits_{i_2=1}^{i-1}\frac{\sgn(m_2)^{i_2+1}}{i^{|m_2|}}\nonumber}&&\\
&=&\left\{\begin{array}{lr}\displaystyle\frac{1}{i^{m_1}}\sum\limits_{i_2=1}^{i-1}\frac{1}{i^{m_2}}\qquad\qquad \qquad\qquad \qquad & m_2>0\\ & \\
\displaystyle\frac{1}{i^{m_1}}\sum\limits_{i_2=1}^{i-1}\frac{(-1)^{i_2+1}}{i^{|m_2|}}=-\frac{1}{i^{m_1}}\sum\limits_{i_2=1}^{i-1}\frac{(-1)^{i_2}}{i^{|m_2|}} & m_2<0\,\end{array}\right.
\end{eqnarray}
which is the expected result. For $m_1$ negative one has
\begin{eqnarray}
\lefteqn{Z_i(m_1,m_2)=\frac{(-1)^{i}}{i^{|m_1|}}\sum\limits_{i_2=1}^{i-1}Z_{i_2}(m_2)=\frac{(-1)^i}{i^{|m_1|}}\sum\limits_{i_2=1}^{i-1}\frac{\sgn(m_2)^{i_2+1}}{i^{|m_2|}}\nonumber}&&\\
&=&\left\{\begin{array}{lr}\displaystyle\frac{(-1)^{i}}{i^{|m_1|}}\sum\limits_{i_2=1}^{i-1}\frac{1}{i^{m_2}}\qquad\qquad \qquad\qquad \qquad & m_2< 0\\ & \\
\displaystyle\frac{(-1)^{i}}{i^{|m_1|}}\sum\limits_{i_2=1}^{i-1}\frac{(-1)^{i_2+1}}{i^{|m_2|}}=-\frac{(-1)^{i}}{i^{|m_1|}}\sum\limits_{i_2=1}^{i-1}\frac{(-1)^{i_2}}{i^{|m_2|}} & m_2>0\end{array}\right.
\end{eqnarray}
We turn now to the general case $k>2$. For $m_1$ positive, one has
\begin{eqnarray}
\lefteqn{Z_i(m_{1,\dots,k})=\frac{1}{i^{m_1}}\sum\limits_{i_2=1}^{i-1}Z_{i_2}(m_{2,\dots,k})}&&\nonumber\\
&=&\frac{1}{i^{m_1}}N(m_{2,\dots,k})\sum\limits_{i_2=1}^{i-1}\dots\sum\limits_{i_k}^{i_{k-1}-1}\frac{\sgn(m_2)^{i_2}}{i_2^{|m_2|}}\frac{\sgn(m_2m_3)^{i_3}}{i_3^{|m_3|}}\dots\frac{\sgn(m_{k-1} m_k)^{i_k}}{i_k^{|m_k|}}\nonumber\\
&=&N(m_{1,\dots,k})\frac{1}{i^{ m_1}}\sum\limits_{i_2=1}^{i-1}\dots\sum\limits_{i_k}^{i_{k-1}-1}\frac{\sgn(m_1 m_2)^{i_2}}{i_2^{|m_2|}}\dots\frac{\sgn(m_{k-1} m_k)^{i_k}}{i_k^{|m_k|}},
\end{eqnarray}  
and for $m_1<0$
\begin{eqnarray}
\lefteqn{Z_i(m_{1,\dots,k})=\frac{(-1)^i}{i^{|m_1|}}\sum\limits_{i_2=1}^{i-1}(-1)^{i_2+1}Z_{i_2}(m_{2,\dots,k})}&&\nonumber\\
&=&\frac{\sgn(m_1)^{i}}{i^{|m_1|}}\sum\limits_{i_2=1}^{i-1}(-1)^{k}Z_{i_2}(-m_{2,\dots,k})\nonumber\\
&=&N(m_{2,\dots,k})(-1)^k\frac{\sgn(m_1)^i}{i^{|m_1|}}\sum\limits_{i_2=1}^{i-1}\dots\sum\limits_{i_k}^{i_{k-1}-1}\nonumber\\
&&\qquad\qquad\frac{-\sgn(m_2)^{i_2}}{i_2^{| m_2|}}\frac{\sgn((-m_2)(-m_3))^{i_2}}{i_2^{|m_2|}}\dots\frac{\sgn((-m_{k-1})(- m_k))^{i_k}}{i_k^{|m_k|}}\nonumber\\
&=&\!\!\!N(m_{1,\dots,k})\frac{\sgn(m_1)^i}{i^{|m_1|}}\sum\limits_{i_2=1}^{i-1}\!\!\dots\!\!\sum\limits_{i_k}^{i_{k-1}-1}\frac{\sgn(m_1 m_2)^{i_2}}{i_2^{| m_2|}}\dots\frac{\sgn( m_{k-1}m_k)^{i_k}}{i_k^{|m_k|}},2
\end{eqnarray}
since 
\begin{eqnarray}
(-1)^k N(-m_{2,\dots,k})&=&(-1)^{-k}(-1)^{k-1-\#n(m_{2,\dots,k})}\nonumber\\
&=&(-1)^{-\#n(m_{2,\dots,k}-1)} =(-1)^{\#n(m_{1,\dots,k})} 
\end{eqnarray}
The connection between the HPL and the MZV is easy to figure out from the definition (\ref{MZVdef}) of the MZV. One obtains the result (\ref{HPLMZVlink})
\begin{equation}
H(\{m_{1,\dots,k}\},1)=N(m_{1,\dots,k})\zeta(\tilde m_{1,\dots,k}).
\end{equation}
\section{Table of representation through more common functions}\label{identities}
\subsection{Weight 2}
\begin{eqnarray}
H\left(\{1, 1\}; x\right) &=& \frac{1}{2} \log\left(1 - x\right)^2 \\
H\left(\{1,-1\};x\right)&=&
  Li_{2}((1-x)/2)-\log(2) \log(1-x)-Li_{2}\left(\frac{1}{2}\right)\\
H\left(\{-1,1\};x\right)&=&
  Li_{2}((1+x)/2)-\log(2) \log(1+x)-
      Li_{2}\left(\frac{1}{2}\right)\\
H\left(\{-1,0\};x\right)&=&
  \log(1+x) \log(x)+Li_{2}(-x)\\
H\left(\{-2\};x\right)&=&-Li_{2}(-x)
\end{eqnarray}
\subsection{Weight 3}
\begin{eqnarray}
H\left(\{1, 2\}; x\right) &=& -2 S_{1,2}\left( x\right) -\log\left(1 - x\right) Li_2\left( x\right)  \\
H\left(\{1, 1, 1\};x\right) &=& -\frac{1}{6} \log\left(1 - x\right)^3  \\
H\left(\{2, 1\}; x\right) &=& S_{1,2}\left( x\right)  \\
H\left(\{2,-1\};x\right)&=&  Li_{3}(\frac{2x}{1+x})-Li_{3}\left(\frac{x}{1+x}\right)-Li_{3}\left(\frac{1+x}{2}\right)\nonumber\\
&&-Li_{3}(x)+\log(1+x) Li_{2}(\frac{1}{2})+\log(1+x) Li_{2}(x)\nonumber\\
&&+\frac{1}{2}  \log(2) \log(1+x)^2+Li_{3}\left(\frac{1}{2}\right)\\
H\left(\{-2,1\};x\right)&=&-S_{1,2}(x)+Li_{3}\left(\frac{-2 x}{1-x}\right)-Li_{3}\left(\frac{1-x}{2}\right)-Li_{3}(-x)\nonumber\\
&&+Li_{3}\left(\frac{1}{2}\right)+Li_{3}(x)+\log(1-x) Li_{2}(-x)\nonumber\\
&&+\log(1-x) Li_{2}\left(\frac{1}{2}\right)-\log(1-x) Li_{2}(x)\nonumber\\
&&+\frac{1}{2} \log(2) \log(1-x)^2-1/6 \log(1-x)^3\\
H\left(\{-2,-1\};x\right)&=&S_{1,2}(-x)\\
H\left(\{1,-1,-1\};x\right)&=&-\frac{1}{2} \log\left(\frac{1-x}{2}\right) \log(1+x)^2-Li_{3}\left(\frac{1}{2}\right)\nonumber\\
&&-\log(1+x) Li_{2}\left(\frac{1+x}{2}\right)+Li_{3}\left(\frac{1+x}{2}\right)\\
H\left(\{-1,1,1\};x\right)&=&  \frac{1}{2} \log\left(\frac{1+x}{2}\right) \log(1-x)^2+Li_{3}\left(\frac{1}{2}\right)\\
&&+\log(1-x) Li_{2}\left(\frac{1-x}{2}\right)-      Li_{3}\left(\frac{1-x}{2}\right)\\
H\left(\{-1,-2\};x\right)&=&  H\left(\{-2\};x\right) H\left(\{-1\};x\right)-2  H\left(\{-2,-1\};x\right)\\
H\left(\{-1,2\};x\right)&=&  H\left(\{-1\};x\right) H\left(\{2\};x\right)-H\left(\{-2,1\};x\right)\nonumber\\
&&-H\left(\{2,-1\};x\right)\\
H\left(\{-1,1,-1\};x\right)&=&  H\left(\{-1\};x\right) H\left(\{1,-1\};x\right)-2  H\left(\{1,-1,-1\};x\right) 
\end{eqnarray}
\subsection{Weight 4}
\begin{eqnarray}
H\left(\{1, 3\}; 
    x\right) &=& -\frac{1}{2}  Li_2\left(x\right)^2 - 
      \log\left(1 - x\right) Li_3\left( x\right)  \\
H\left(\{1, 1, 2\}; x\right) &=& 
  \frac{1}{2} \log\left(1 - x\right)^2 Li_2\left( x\right) + 2 \log\left(1 - x\right) S_{1,2}\left( x\right) \nonumber\\
  &&+ 
      3 S_{1,3}\left( x\right)  \\
H\left(\{1, 2, 1\}; 
    x\right) &=& -\log\left(1 - x\right) S_{1,2}\left( x\right) - 
      3 S_{1,3}\left( x\right)  \\
H\left(\{2, 2\}; 
    x\right) &=& -2 S_{2,2}\left( x\right) + 
      \frac{1}{2} Li_2\left(x\right)^2 \\  
H\left(\{3, 1\}; x\right) &=& S_{2,2}\left( x\right)\\  
H\left(\{2, 1, 1\}; x\right) &=& 
  S_{1,3}\left( x\right)\\  
H\left(\{1, 1, 1, 1\}; x\right) &=& 
  \frac{1}{24} \log\left(1 - x\right)^4  
\end{eqnarray}
\subsection{Arbitrary weight}
\begin{eqnarray}
H\left(\{n\}; x\right) &=& 
  Li_n\left(x\right),\qquad  n>0\\
 H\left(\{n\}; x\right) &=& 
  -Li_{-n}\left(-x\right),\qquad  n<0\\   
H(\{^n 1\}; 
    x) &=& (-1)^n \frac{\log\left(1 - x\right)^n}{n!}  \\
H(\{^n(-1)\}; 
    x) &=& \frac{\log\left(1+x\right)^n}{n!}  \\
H(\{n,^{p-1}1\}; x) &=& 
  S_{n-1,p}\left( x\right)  
\end{eqnarray}


\end{document}